\documentclass[a4paper,12pt]{article}
\usepackage[utf8]{inputenc}
\usepackage{cancel}
\usepackage{ulem}
\usepackage{amsfonts}
\usepackage{amssymb}
\usepackage{graphicx}
\usepackage{amsmath}
\usepackage{enumerate}
\usepackage{here}
\usepackage{placeins}
\usepackage{subfig}
\usepackage{color}
\usepackage{float}
\usepackage{here}
\usepackage{cite}
\usepackage{mathrsfs}
\usepackage{float,epsfig}
\usepackage{dcolumn}
\usepackage{subfloat}
\usepackage{graphicx}
\usepackage{bm}
\usepackage{amsmath,amssymb,amsthm}
\usepackage[colorlinks=true,linkcolor=blue]{hyperref}
\hypersetup{citecolor=magenta}
\title{\bf \Large  Phase transitions and geothermodynamics of black holes in dRGT massive gravity}

\author{M. Chabab$^{1}$\footnote{mchabab@uca.ac.ma (Corresponding author)}, H. El Moumni$^{1,2}$\footnote{hasan.elmoumni@edu.uca.ma}, S. Iraoui$^{1}$\footnote{samir.iraoui@ced.uca.ma}, K. Masmar$^{1}$\footnote{k.masmar@edu.uca.ac.ma} \\
	\\ 
	{\small $^{1}$ High Energy and Astrophysics Laboratory,  Faculty of Science Semlalia,}\\
         { \small Cadi Ayyad University, P.O.B. 2390 Marrakech, Morocco.
	}\\
	{\small $^{2}$ EPTHE, Physics Department, Faculty of Science,  Ibn Zohr University, Agadir, Morocco. }
}

\date{}
\begin{document} \maketitle
	
\begin{abstract}
 In this paper, we study the thermodynamics and geothermodynamics of spherical black hole solutions in dRGT massive gravity in a new extended phase space. Inspired by the work of Kastor et al. \cite{Kastor}, by interpreting the graviton mass as a thermodynamical variable, we propose a first law of thermodynamics which  include a mass term and establish a new Smarr Formula. Then, we perform a thermodynamical analysis to reveal the existence of a critical behavior for black holes in dRGT massive gravity with two different critical points through canonical and grand canonical ensembles. To consolidate these results, we make use of the thermodynamical geometry formalism, with the  {\tt HPEM}  and the Gibbs free energy metrics, to derive the singularities of Ricci scalar curvatures and show that they coincide with those of the capacities. The effect of different values of the spacetime parameters on the stability conditions is also discussed. 
\end{abstract}

\tableofcontents

\section{Introduction}
 In Einstein' theory of general relativity, the graviton is a massless particle. One might raise the natural and legitimate question of whether a self-consistent gravity theory with massive graviton is achievable. A massive gravity came into existence as a straightforward modification of the general relativity by considering consistent interaction terms that can be interpreted as a graviton mass.

Massive gravity theories have become increasingly popular in the current literature. The first construction of a linear theory of massive gravity with a ghost-free theory of non-interacting massive gravitons goes back to Fierz and Pauli' work $(FP)$  \cite{Fierz1}. By adding interaction terms in the linearized level of general relativity, FP showed that the mass term must be of the form $ m_{g}^{2}\left(h^{2}-h_{\mu\nu}h^{\mu\nu}\right)$. Unfortunately, the theory of Fierz and Pauli suffered from a discontinuity problem in the zero mass limit and which prevent recovery of the general relativity results at the low energy scales. This is called van Dam-Veltman-Zakharov (vDVZ) discontinuity \cite{vanDam:,Zakharov}.

In addition, a ghost propagating degree of freedom, dubbed Boulware-Deser ghost $(BD)$ \cite{Boulware,Boulware2}, shows up at the nonlinear level in a generic  FP theory. This issue has been resolved recently by de Rham, Gabadadze and Tolley in the framework of dRGT massive gravity \cite{deRham}, who found a two-parameter family of nonlinear theories free from the BD ghost order-by-order and to all orders, at least in the decoupling limit.  Then, de Rham et al.  proposed a candidate theory of massive gravity free of BD ghost \cite{deRhamdRGT}.  The latter is automatically ghost free  to all orders in the decoupling limit with the Hamiltonian constraint maintained at least up to quartic order away from the decoupling limit. Later, a generalization to a ghost free non-linear massive gravity action for all orders was proposed in \cite{HassanF}. Until now, a four-dimensional black hole solution with a Ricci flat horizon has been constructed in \cite{Vegh}. The exact spherical black hole solution in de dRGT massive gravity used in this paper has been obtained in  \cite{Ghosh}.\\

 The graviton mass in massive gravity naturally generates the cosmological constant and the global monopole term. Hence such theory can provide the solution for describing our universe which is currently expanding with acceleration without introducing any dark energy or cosmological constant \cite{Vasiliev:1995dn,Abbott}. Current experimental observations put constraints on the mass of the graviton $m_{g}$, particularly from the recent detection  of gravitational waves by advanced LIGO, which provided an upper limit : $m_{g}<1.2\times 10^{-22} eV/c^{2}$ \cite{Abbott}.

Recently, the study of phase transitions of black holes in asymptotically Anti de-Sitter spacetime has aroused growing interest \cite{Kubiznak13,Belhaj2,Belhaj1} since these transitions have been related with holographic superconductivity in the context of the AdS/CFT correspondence \cite{Maldacena:1997re,Hartnoll:2008kx,holography,Zeng:2014uoa}. In this respect, many attempts have been made to establish relationship between the thermodynamic phase transitions  and quasinormal modes of black holes \cite{Liu,Chabab1,Chabab2}.  The phase transitions are usually signaled by the existence of  discontinuities of a state space variable,  via either the analysis of the heat capacity \cite{Belhaj1} or by using the geothermodynamics approaches \cite{ Chabab3,Chabab4, Liu2010sz} proposed  by Ruppeiner \cite{hep49}, Quevedo \cite{hep54},  HPEM \cite{Newmetric} or by Liu-Lu-Luo-Shao \cite{Liu2010sz}. \\

Hence, our aim in this paper is to analyze the extended phase structure and investigate the critical behavior of  both neutral and charged black holes in dRGT massive gravity by using the geometrical studies. \\

This paper is organized as follows: In the next section, we present a brief review of the thermodynamics of dRGT massive gravity black holes. In section \ref{sec3}, by considering the graviton mass as a thermodynamic variable, we explore the critical behavior and the phase structure for neutral black holes, either in the canonical ensemble by fixing the graviton mass and in the grand canonical ensemble where the graviton potential is a constant. In section \ref{sec4}, we extend the analyses of section  \ref{sec3}  to the charged black holes.
In section \ref{sec5}, we investigate the criticality from the geothermodynamical point of view with the $HPEM$  and the Gibbs free energy metrics. The last section is devoted to our conclusion.
\FloatBarrier

\section{Massive gravity background and thermodynamics}\label{sec2}
\subsection{General background}
The massive gravity model considered in this paper can be described by the following action  \cite{deRhamdRGT}
\begin{equation}\label{action}
S=\frac{1}{2\kappa} \int d^4 x \sqrt{-g}\left[R+ m_g^2 \mathcal{U}(g,\phi^2)\right],
\end{equation}
the first term is the scalar curvature and ${\cal U}$ is an interaction mass term corresponding to the massive graviton. The latter modifies the gravitational sector with the introduction the parameter $m_g$ interpreted as  graviton mass. The effective potential ${\cal U}$ in four-dimensional spacetime is given by,
\begin{eqnarray}\label{potential}
 {\cal U} = {\cal U}_2 + \alpha_3{\cal U}_3 +\alpha_4{\cal U}_4 ,
\end{eqnarray}
where $\alpha_3$ and $\alpha_4$ are dimensionless free parameters of the theory. We can write the interaction in terms of the tensor ${\cal K}^\mu_\nu$  defined as;
\begin{eqnarray}
{\cal K}^\mu_\nu =
\delta^\mu_\nu-\sqrt{f_{ab}
	\partial^{\mu}\phi^a\partial_\nu\phi^b}. \label{K-tensor}
\end{eqnarray}
Here the square root is understood as $\left(\sqrt{\mathcal{A}}\right)^{\mu}_{\sigma}\left(\sqrt{\mathcal{A}}\right)^{\sigma}_{\nu}=\mathcal{A}_{\nu}^{\mu}$, while $f_{ab}$ is a reference  metric (fixed symmetric tensor) and
\begin{eqnarray}
 {\cal U}_2&\equiv&[{\cal K}]^2-[{\cal K}^2] ,\\
 {\cal U}_3&\equiv&[{\cal K}]^3-3[{\cal K}][{\cal K}^2]+2[{\cal K}^3] ,\\
 {\cal U}_4&\equiv&[{\cal K}]^4-6[{\cal K}]^2[{\cal K}^2]+8[{\cal K}][{\cal
K}^3]+3[{\cal K}^2]^2-6[{\cal K}^4],
\end{eqnarray}
 $[{\cal K}]$ denotes the traces,
 $[{\cal K}]=Tr({\cal K})={\cal K}^\mu_\mu$ and $[{\cal K}^n]=({\cal K}^n)^\mu_\mu$. The four scalar fields $\phi^a$ are the St\"uckelberg scalar fields introduced to restore general covariance of the theory.

Following the convention of  \cite{Ghosh},  we introduce the new variable $\alpha$ and $\beta$ as
\begin{equation}
\alpha_3=\frac{\alpha-1}{3},\quad \alpha_4=\frac{\beta}{4}+\frac{1-\alpha}{12}.
\end{equation}
By considering the reference metric $f_{ab}=\left(0,0,c^{2},c^{2} \sin^{2}\theta\right)$ \cite{Cai}, the static and spherically symmetric black hole solution of Eq.~\eqref{action} reads as  \cite{Ghosh},
\begin{equation}
ds^2= -f(r)dt^2+\frac{dr^2}{f(r)}+r^2 d\Omega^2,
\end{equation}
here, the radial function $f(r)$ reads as,
\begin{equation}\label{fr1}
	f(r)=1 -\frac{2 M}{r}+\frac{Q^2}{r^2}-\frac{\Lambda  }{3} r^2+\gamma  r+\zeta,
\end{equation}
with
\begin{subequations}
\begin{eqnarray}
\Lambda&=&-3 m^2_g \left(1+\alpha+\beta\right),\\
\gamma&=&-c m^2_g \left(1+2\alpha+3\beta\right),\\
\zeta&=&c^2 m^2_g \left(\alpha+3\beta\right),
\end{eqnarray} \label{pardef}
\end{subequations} 
where $c$ is an arbitrary parameter, $M$ an integration constant related to the mass of the black hole. We have not considered an  AdS or dS space in the problem, but the graviton mass $m_{g}$ produces an identifiable term with a nonzero cosmological constant ($1+\alpha+\beta \ne 0$). In this study we consider AdS space, where we impose the condition $\alpha+\beta+1>0$. We note that at the limit $\alpha=-1$ and $\beta=1/3$, we recover a similar of RN-AdS black hole where $-m_g^2$ play the role of the cosmological constant.

\subsection{First law and Smarr formula}
 Kastor et al. have interpreted $\Lambda$ as a thermodynamic variable and defined thermodynamically conjugate variable as the thermodynamic volume \cite{Kastor}. Here, we extend this idea by treating the graviton's mass as a thermodynamic variable while its conjugate is identified with the  graviton potential. Then,  we perform a study of the thermodynamics and geothermodynamics of the neutral and charged black hole  in dRGT massive gravity background.

The event horizon $r_h$ of the black hole is determined by the equation $\left.f(r)\right|_{r=r_{h}} = 0$. Then, according to Eq.~\eqref{fr1}, the mass of the black hole can be expressed as:
\begin{equation}\label{mass}
M= \frac{3 Q^2-\Lambda  r_{h}^4+3 r_{h}^2 (\zeta +\gamma  r_{h}+1)}{6 r_{h}}.
\end{equation}
The black hole entropy  may be determined from the Bekenstein-Hawking formula \cite{Bekenstein,Hawking}
\begin{equation}
S=\pi r_h^2,
\end{equation}
therefore, by using Eq.~\eqref{pardef} and Eq.~\eqref{mass}, the mass of the black hole is re-written as a function of $m_g$ and $S$,
\begin{equation}\label{massexp}
M=\frac{S m_g^2 \left(\pi  c^2 \alpha '-\sqrt{\pi } c \sqrt{S} \beta '+S \gamma
	'\right)+\pi  \left(\pi  Q^2+S\right)}{2 \pi ^{3/2} \sqrt{S}}.
\end{equation}
Here, to simplify the expressions, we use the notations:
$$\alpha^\prime=\alpha+3\beta,$$
$$\beta^\prime=1+2\alpha+3\beta,$$
$$\gamma^\prime=1+\alpha+\beta.$$ 
The temperature reads as, 
\begin{equation}\label{temperature}
T=\left.\frac{\partial M}{\partial S}\right|_{m_g,Q}=\frac{S m_g^2 \left(\pi  c^2 \alpha '-2 \sqrt{\pi } c \sqrt{S} \beta '+3 S \gamma
	'\right)+\pi  \left(S-\pi  Q^2\right)}{4 \pi ^{3/2} S^{3/2}},
\end{equation}
which is precisely the Hawking temperature of the black hole \footnote{
$T=\frac{\kappa}{2\pi}=\left.\frac{f'\left(r\right)}{4\pi}\right|_{r=r_{h}}=\frac{S m_g^2 \left(\pi  c^2 \alpha '-2 \sqrt{\pi } c \sqrt{S} \beta '+3 S \gamma
	'\right)+\pi  \left(S-\pi  Q^2\right)}{4 \pi ^{3/2} S^{3/2}}$ (see S.~G.~Ghosh and S.~D.~Maharaj, PRD 89 (2014) 084027)}.

Note here that, one recovers the temperature of RN black holes in the massless limit $m_g=0$. Moreover, one can also derive the static electric potential $\Phi$ associated with the charge of the black hole as well as the graviton potential $\mu$ conjugate to the graviton mass as,
\begin{eqnarray}\label{chimicalpot}
\Phi&= &\left.\frac{\partial M}{\partial Q}\right|_{S, m_g}=\sqrt{\pi} Q S^{-\frac{1}{2}},\\
\label{chimique}
\mu&=&\frac{\sqrt{S} m_g \left(\pi  c^2 \alpha '-\sqrt{\pi } c \sqrt{S} \beta '+S \gamma
	'\right)}{\pi ^{3/2}}.
\end{eqnarray}
From Eq.~\eqref{chimique}, we see clearly that the graviton potential $\mu$ is not sensitive to the charge of the black hole.

To ensure the Smarr relation, we use the recipe suggested in \cite{Ning} and assume that $\alpha^\prime$ and $\beta^\prime$ as variables, 
	\begin{equation}\label{Smart}
	M=2 T S+Q \Phi- \mu m_g +2
	\mu _{\alpha^\prime} 	\alpha^\prime+ \mu _{\beta^\prime} \beta^\prime,
	\end{equation}
 where $\mu_{\alpha^\prime}$ and $\mu_{\beta^\prime}$  are the conjugate quantities of $\alpha^\prime$ and $\beta^\prime$ respectively  \footnote{Note here that if $c$ is introduced as thermodynamical variable, then the Smarr formula becomes, 
$M=2 T S+Q \Phi- \mu m_g +2\left(1-\lambda\right)\mu _{\alpha^\prime} 	\alpha^\prime+ \left(1-\lambda\right)\mu _{\beta^\prime} \beta^\prime + \lambda\mu _{c} c$, 
where $\lambda$ is an arbitrary constant. We can thus  see that the contribution of $\alpha'$ and $\beta'$ are lost when $\lambda=1$, a scenario that we would like to avoid.} 
For the similar RN-AdS BH ($\alpha=-1$ and $\beta=1/3$) we get the formula,\cite{Mann}
	$$M=2TS+Q\Phi-2 PV,$$ 
	where $\Lambda$ is identified with a negative pressure $P=-\frac{\Lambda}{8 \pi}=\frac{m_g^2}{8 \pi}$ and $V=\frac{\partial M}{\partial P}=\frac{4 }{3} \pi r_{h}^{3}$  its thermodynamical conjugate.
 \FloatBarrier
\section{Thermodynamic  behaviour of neutral black holes}\label{sec3}
In this section, we study the thermodynamic phase transitions by considering the black hole either as a closed system (canonical ensemble), or as an open system (grand canonical ensemble).
\subsection {The canonical ensemble}
\begin{figure}[!h]
	\begin{center}
		\includegraphics[width=8cm,height=5cm]{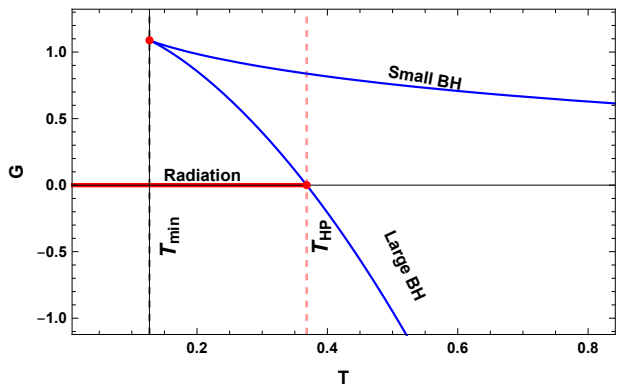}
	\caption{Canonical ensemble: The Gibbs free energy as function of the temperature for $m_g=1,\ c=1,\ \alpha=1\ \text{and}\ \beta=2$.}  \vspace*{-.2cm}
	\label{fig2}
\end{center}
\end{figure}

 For a chargeless system,  the free energy reads as \cite{Mann},
\begin{equation}\label{gibbsneutre}
G=M-T S=\frac{\sqrt{S} \left(m_g^2 \left(\pi  c^2 \alpha '-S \gamma '\right)+\pi
	\right)}{4 \pi ^{3/2}}.
\end{equation}
We plot in Fig.~\ref{fig2},  $G$ with respect to the Hawking temperature for a fixed graviton mass. We have to stress that a direct substitution of the entropy in terms of temperature is non trivial, we thus get around this difficulty and evaluate first the values of $G$ for different values of $S$ and then determine the corresponding values of $T$ from Eq.~\eqref{temperature}. The temperature  shows a minimum given by, 
\begin{equation}\label{tmin}
T_{min}=m_g \dfrac{ \sqrt{3} \sqrt{\gamma^\prime \left(c^2 \alpha^\prime m_g^2+1\right)}-c \beta^\prime m_g}{2 \pi },
\end{equation}
corresponding to the minimal entropy:
\begin{equation}\label{minT}
S_{min}=\pi \dfrac{ c^2 m_g^2 \alpha^\prime+1}{3 m_g^2 \gamma^\prime}.
\end{equation}

Fig.~\ref{fig2} clearly shows several features:  $1)$ No black hole solution cannot exist below the minimal temperature there. $2)$  There exist two branches above $T_{min}$, the upper one is for small black holes which is thermodynamically unstable, while the lower branch corresponds to a thermodynamically stable large black holes. The positive Gibbs free energy for $T<T_{HP}$ shows that the system is in a radiation phase.

Furthermore, at $T_{HP}$,  the intersection point of the bottom branch and $G=0$ indicates the well known Hawking-Page phase transition \cite{Hawking}. We can see from Eq.~\eqref{gibbsneutre} that this point occurs at:
\begin{equation}
S_{HP}=\pi  \dfrac{c^2 m_g^2 \alpha^\prime+1}{m_g^2 \gamma^\prime}.
\end{equation}

After having identified the thermodynamic behaviour of neutral black holes in dRGT massive gravity, it is important to investigate the heat capacity as a state space variable. Afterwards, we will  study the thermodynamic geometry of the neutral black holes with a fixed graviton mass.  Fixing the graviton mass  means that the cosmological constant is kept constant at a certain value and not treated as a thermodynamic variable. 

\begin{figure}[!h]
				\begin{center}
		\includegraphics[width=8cm,height=5cm]{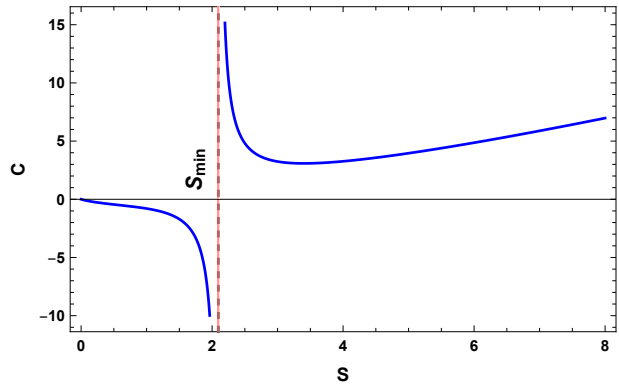}
		\caption{Canonical ensemble: The heat capacity as a function of the entropy for the neutral BH, with $m_g=1,\ c=1,\ \alpha=1\ \text{and}\ \beta=2$.}
		\label{fig71}
	\end{center}
	\end{figure}

Using Eq.~\eqref{temperature}  the heat capacity of the neutral black hole for a fixed $m_g$, given by $C_{m_g}=T\left(\frac{\partial S}{\partial T}\right)_{m_g}$, can be written as

\begin{equation} \label{Heat}
C_{m_g}=2 S \frac{m_g^2 \left(\pi  c^2 \alpha '-2 \sqrt{\pi } c \sqrt{S} \beta '+3 S \gamma
	'\right)+\pi }{m_g^2 \left(3 S \gamma '-\pi  c^2 \alpha '\right)-\pi },
\end{equation}
it can be seen that the heat capacity diverges at $S=\pi  \frac{c^2 m_g^2 \alpha^\prime+1}{3 m_g^2 \gamma^\prime}$, which is just  the point corresponding to the minimal Hawking temperature given by $S_{min}$ in Eq.~\eqref{minT}, of course, we can verify that the numerator of $C_{m_g}$ does not vanish at this point. From the figure \ref{fig71},  we note that the heat capacity is negative for $S < S_{min}$, while it is positive for $S > S_{min}$. A positive heat capacity indicates the thermodynamic stability, whereas, a negative heat capacity corresponds to a thermodynamic instability.

\FloatBarrier
\subsection {The grand canonical ensemble}
In the grand canonical ensemble, the black hole is seen as an open system with a fixed graviton potential $\mu$. In this case, by inverting Eq.~\eqref{chimique}, then substituting in Eq.~\eqref{temperature} and taking the limit of vanishing charge we obtain Hawking temperature:
 
	\begin{equation}
	T=\frac{\pi ^2 \mu ^2 \left(\pi  c^2 \alpha '-2 \sqrt{\pi } c \sqrt{S} \beta '+3 S \gamma
		'\right)+S \left(\pi  c^2 \alpha '-\sqrt{\pi } c \sqrt{S} \beta '+S \gamma
		'\right)^2}{4 \sqrt{\pi } S^{3/2} \left(\pi  c^2 \alpha '-\sqrt{\pi } c \sqrt{S}
		\beta '+S \gamma '\right)^2},
	\end{equation}	
as to the Gibbs free energy, defined as $G=M-T S- \mu m_{g} $,  it can be expressed in the following form:
	\begin{equation}
	G=\frac{\pi ^2 \mu ^2 \left(-3 \pi  c^2 \alpha '+4 \sqrt{\pi } c \sqrt{S} \beta '-5 S
		\gamma '\right)+S \left(\pi  c^2 \alpha '-\sqrt{\pi } c \sqrt{S} \beta '+S \gamma
		'\right)^2}{4 \sqrt{\pi S} \left(\pi  c^2 \alpha '-\sqrt{\pi } c \sqrt{S}
		\beta '+S \gamma '\right)^2}.
	\end{equation}
Here Eq.~\eqref{massexp}, Eq.~\eqref{temperature} and Eq.~\eqref{chimique}  have also been invoked.  

\begin{figure}[!h]
		\begin{center}
\begin{tabbing}
	\hspace{8cm}\=\kill
	\includegraphics[width=7cm,height=4cm]{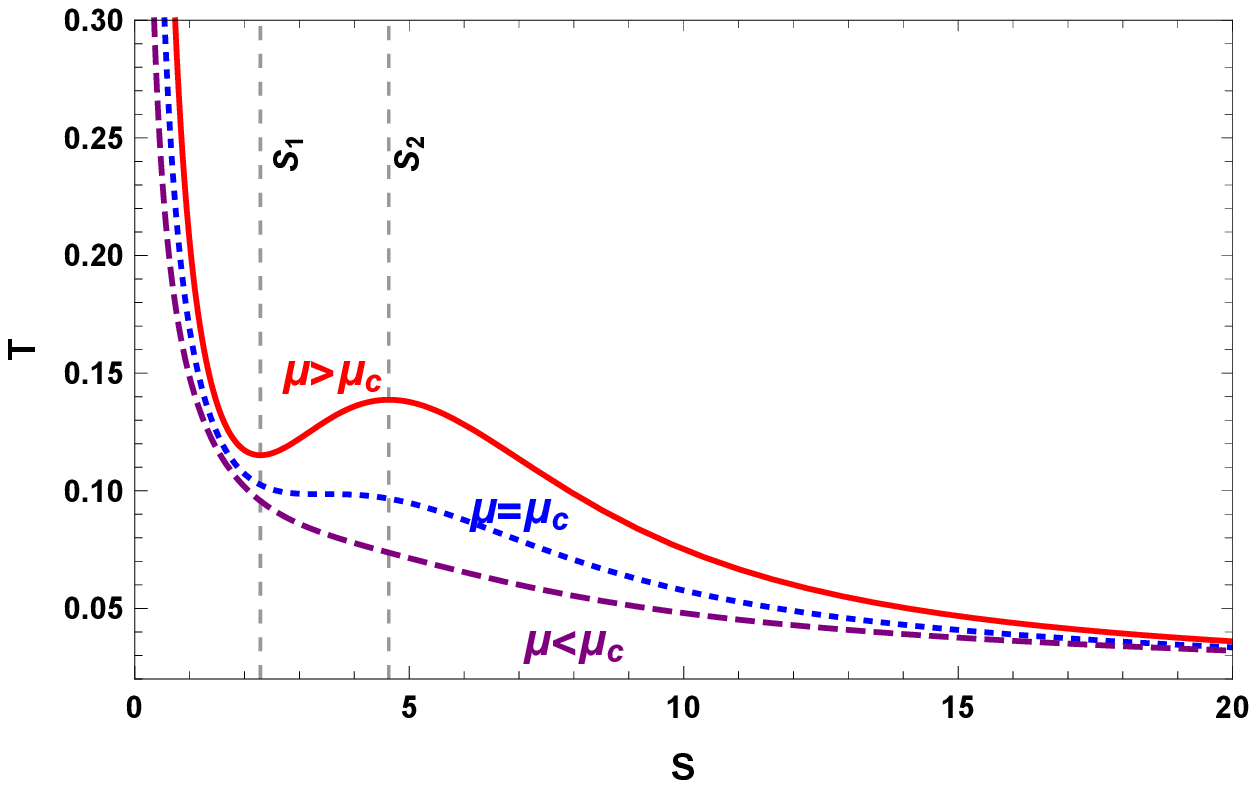} \>
	\includegraphics[width=7cm,height=4cm]{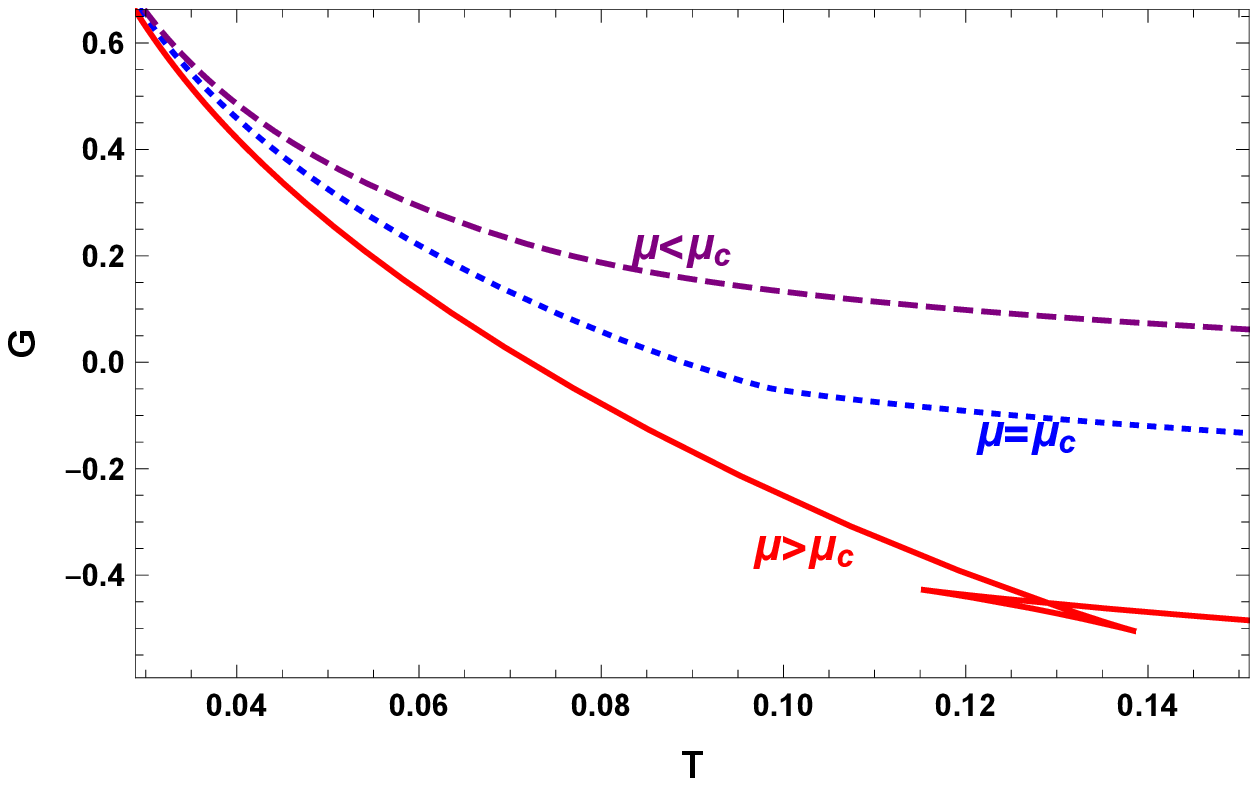}
\end{tabbing} 
	 \caption{Grand canonical ensemble: (left) Temperature as a function of $S$. (right) The Gibbs free energy versus $T$ for the neutral BH. The following values have been used for  $\ c=1,\ \alpha=1\ \text{and}\ \beta=2$.}
	 \label{TGSGCN}
	 \end{center}
\end{figure}

		Compared to  the canonical framework, the thermodynamics behaviour in the grand canonical case, is different. Indeed, we find a critical behaviour and a critical point located at $\mu=\mu_{c}$, while for $\mu>\mu_{c}$,  two critical points $S_{1}$ and $S_{2}$ are revealed, corresponding to the local minimum and maximum respectively, as can be seen from the left panel of figure \ref{TGSGCN}. At $\mu=\mu_{c}$ the two points $S_{1}$ and $S_{2}$ coalesce into a single point located at $S_{c}$ resulting from the conditions
	\begin{equation}\label{critiqueGCN}
	\left.\frac{\partial T}{\partial S}\right|_{\mu_{c},S_{c}}=0,\quad \left.\frac{\partial^2 T}{\partial S^2}\right|_{\mu_{c},S_{c}}=0.
	\end{equation}
	At this stage, it is worth to notice that Eq.~\eqref{critiqueGCN} is difficult to solve analytically. However, 
we can get a numerical solution in a straightforward way by fixing the parameters $\alpha$, $\beta$ and $c$.

\begin{figure}[!h]
 	\begin{center}
	\begin{tabbing}
	\hspace{8cm}\=\kill
	\includegraphics[width=7cm,height=4cm]{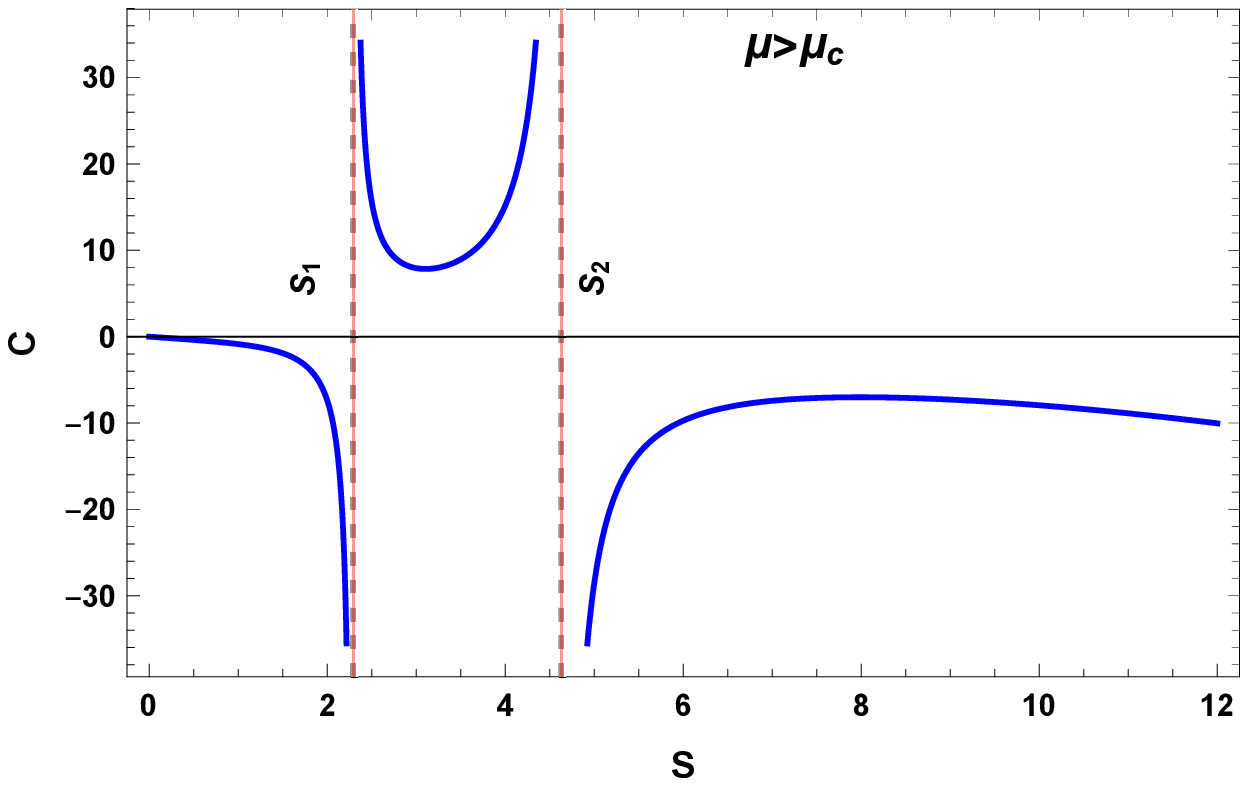} \>
	\includegraphics[width=7cm,height=4cm]{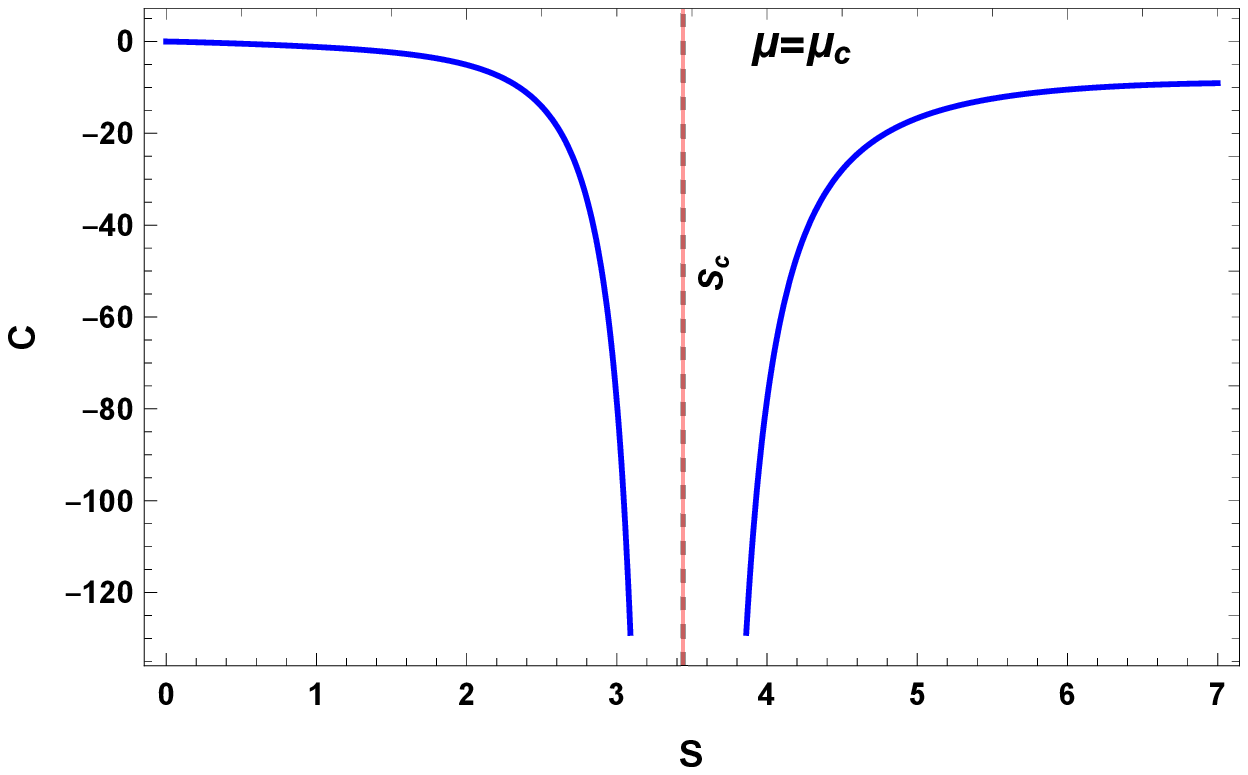}
\end{tabbing} 
	\caption{Grand canonical ensemble: The heat capacity $C_{\mu}$ versus $S$ of neutral BH for: $\mu>\mu_{c}$ (left)  and   $\mu=\mu_{c}$ (right), The dashed lines correspond to the divergency points of $C_{\mu}$;  red lines correspond to critical points. We have set: $\ c=1,\ \alpha=1\ \text{and}\ \beta=2$.} 
	\label{CGCN}
\end{center}
\end{figure}

Again we call upon the heat capacity $C_{\mu}=T\left(\frac{\partial S}{\partial T}\right)_{\mu}$ at constant potential $\mu$ to uncover the local stability in grand canonical ensemble. The behaviour found is reversed compared to that of RN-AdS black hole \cite{Mann}. Hence, we can see from figure \ref{CGCN}, that the locally stable region is located between $S_1$ and $S_2$ where $C_{\mu}$ is positive, while the regions  $S<S_1$ and  $S>S_2$, corresponding respectively to  small and large black holes, are locally instable with negative heat capacity:  $C_{\mu}$ diverges exactly at the critical points.
\FloatBarrier

\section{Thermodynamic behaviour of charged black  holes }\label{sec4}
In this section we aim to discuss the thermodynamic behaviour of charged black hole in the canonical ensemble as well as in the grand canonical ensemble.
\subsection {The canonical ensemble}
\begin{figure}[!h]
		\begin{center}
			\begin{tabbing}
			\hspace{8cm}\=\kill
			\includegraphics[width=7cm,height=4cm]{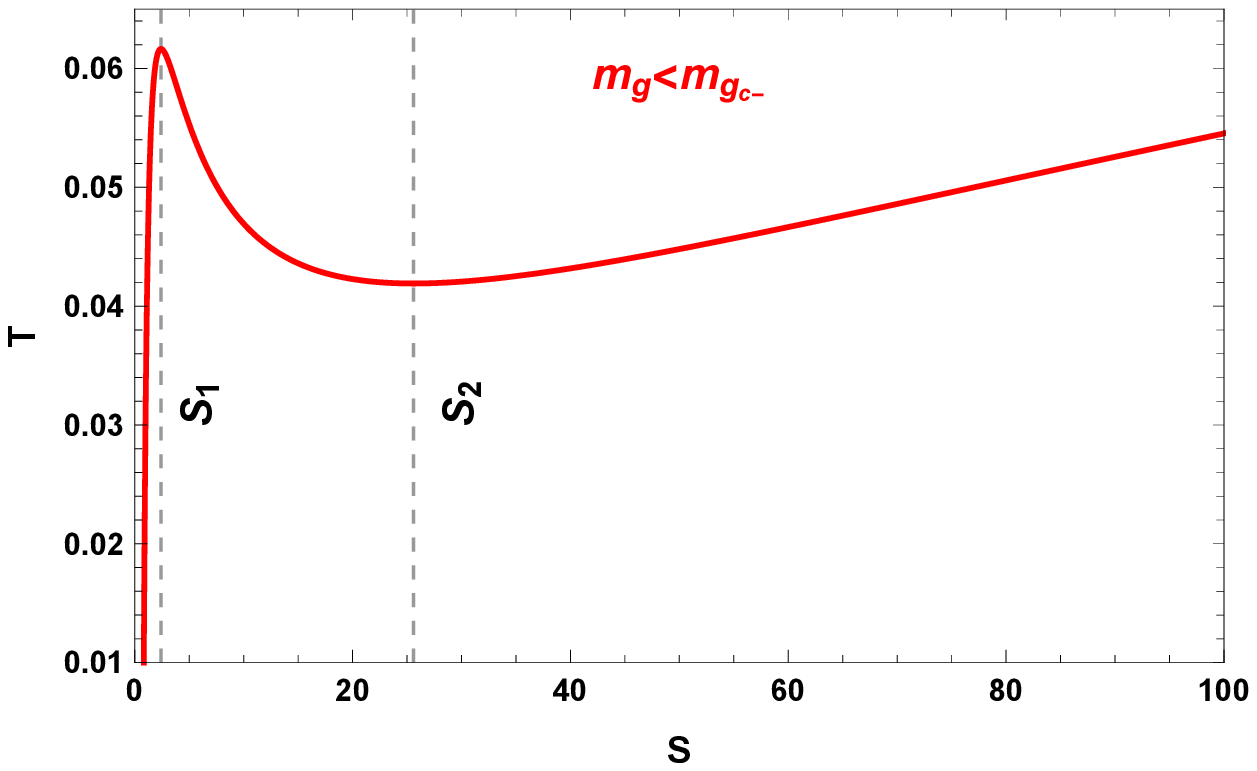} \>
			\includegraphics[width=7cm,height=4cm]{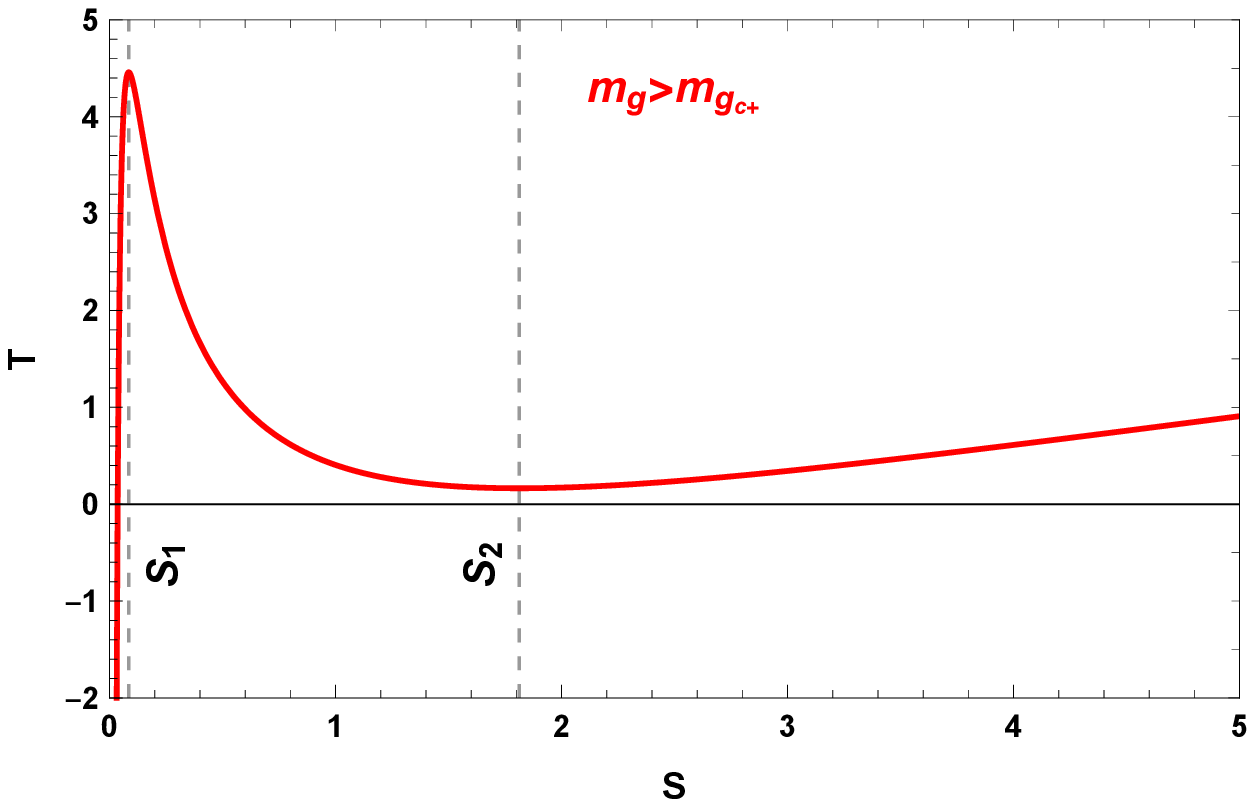} \\
			\includegraphics[width=7cm,height=4cm]{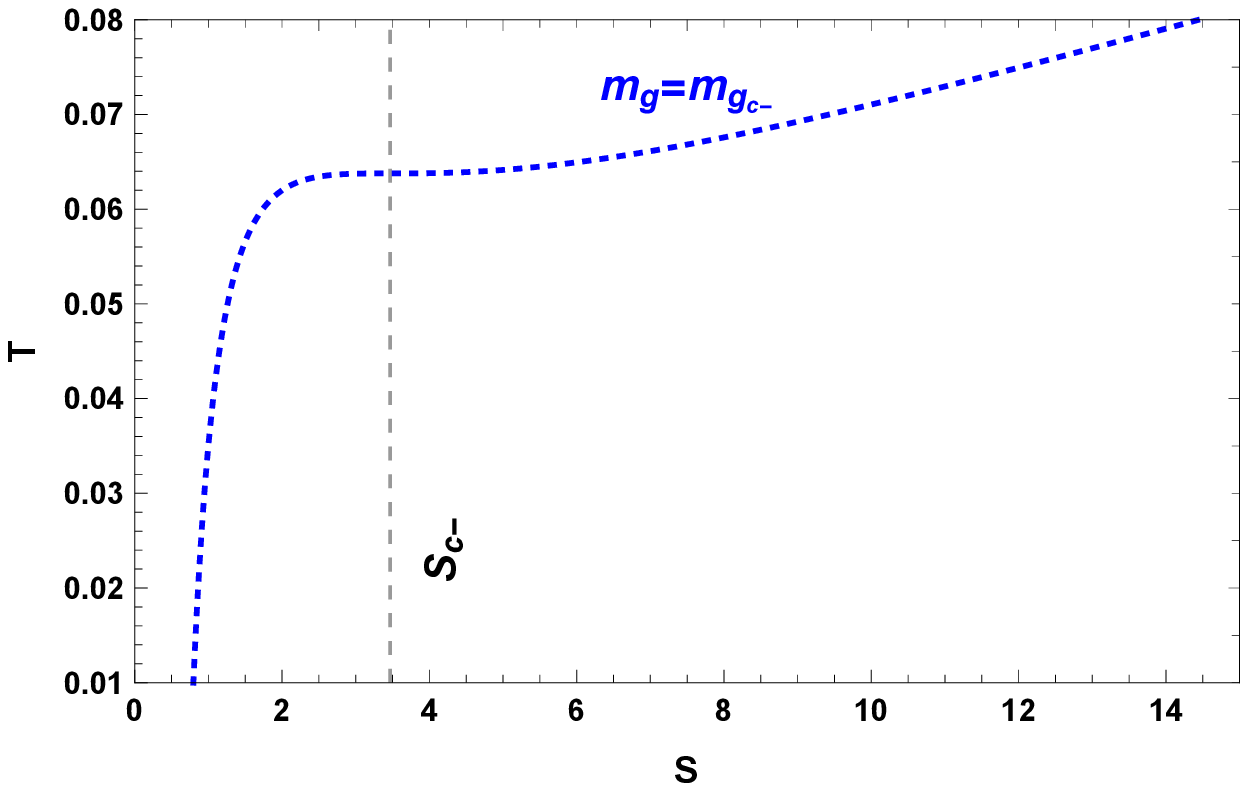} \>
			\includegraphics[width=7cm,height=4cm]{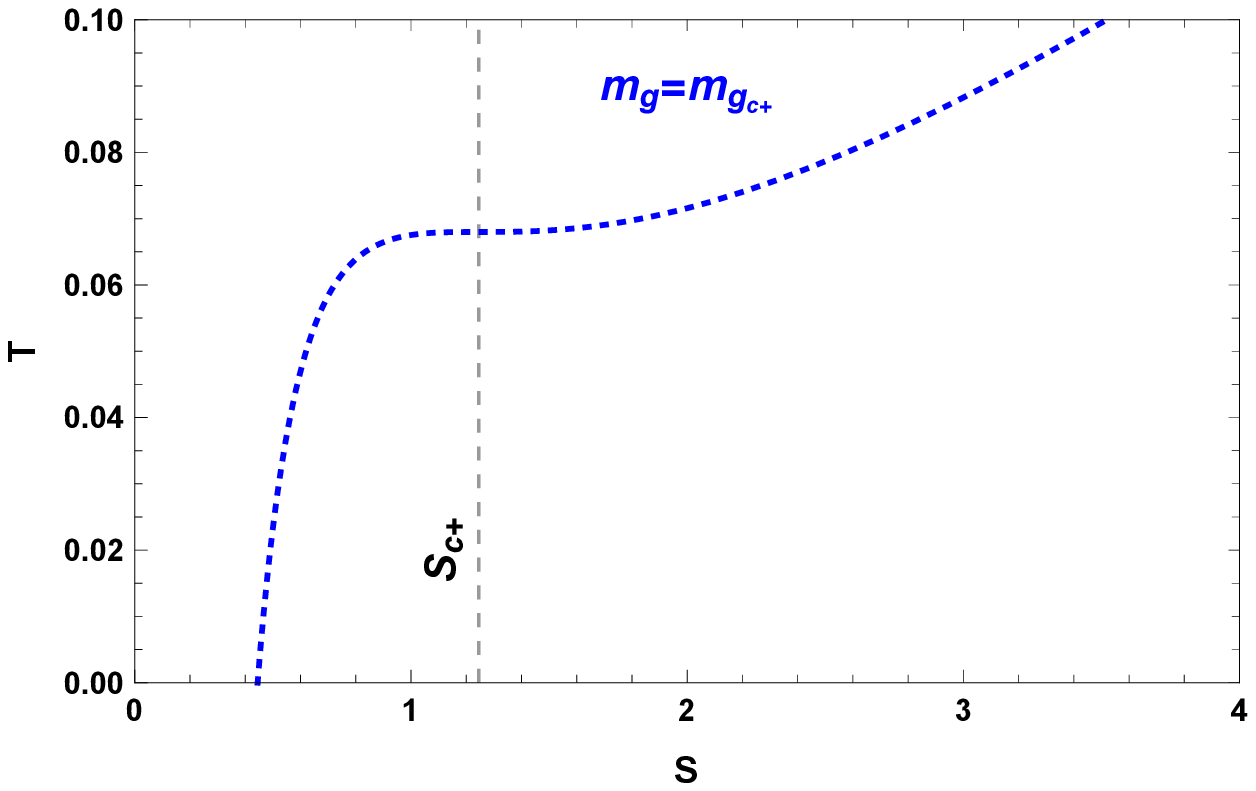} \\
	    \hspace*{4cm}	\includegraphics[width=7cm,height=4cm]{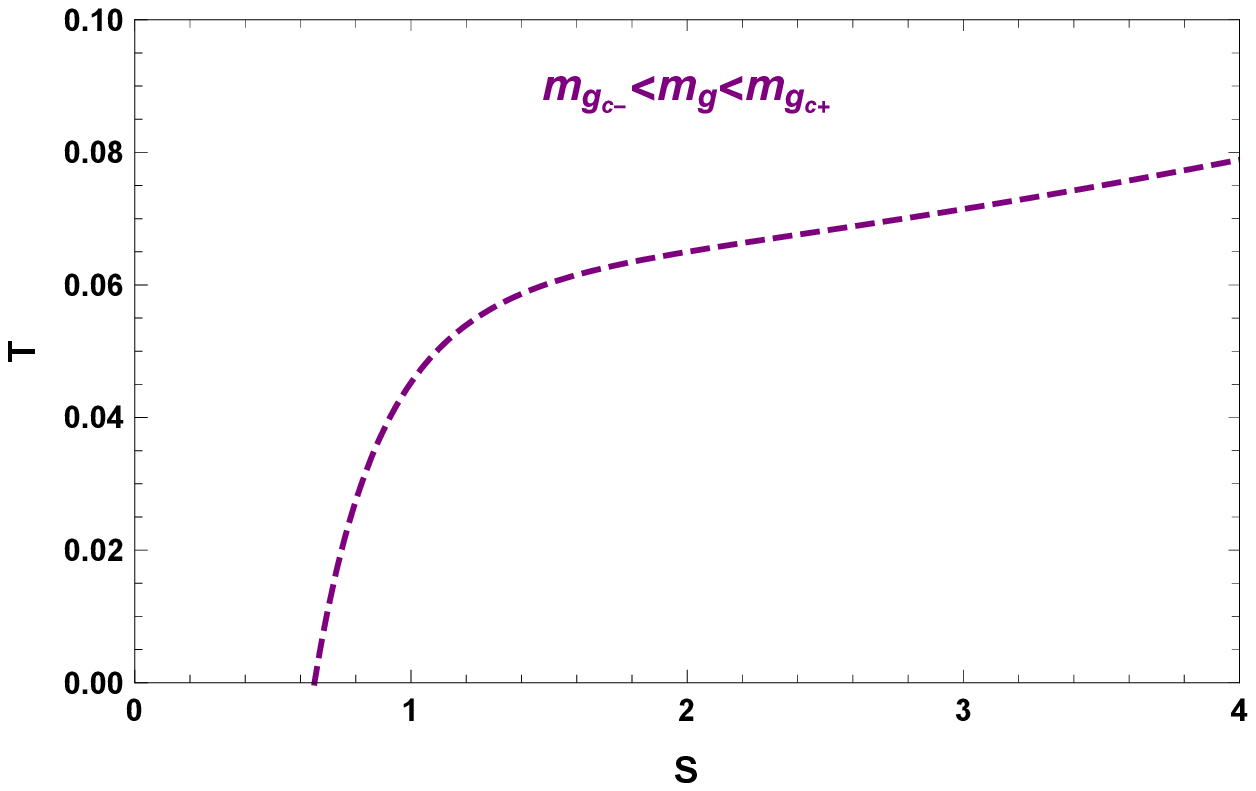} \\
			\includegraphics[width=7cm,height=4cm]{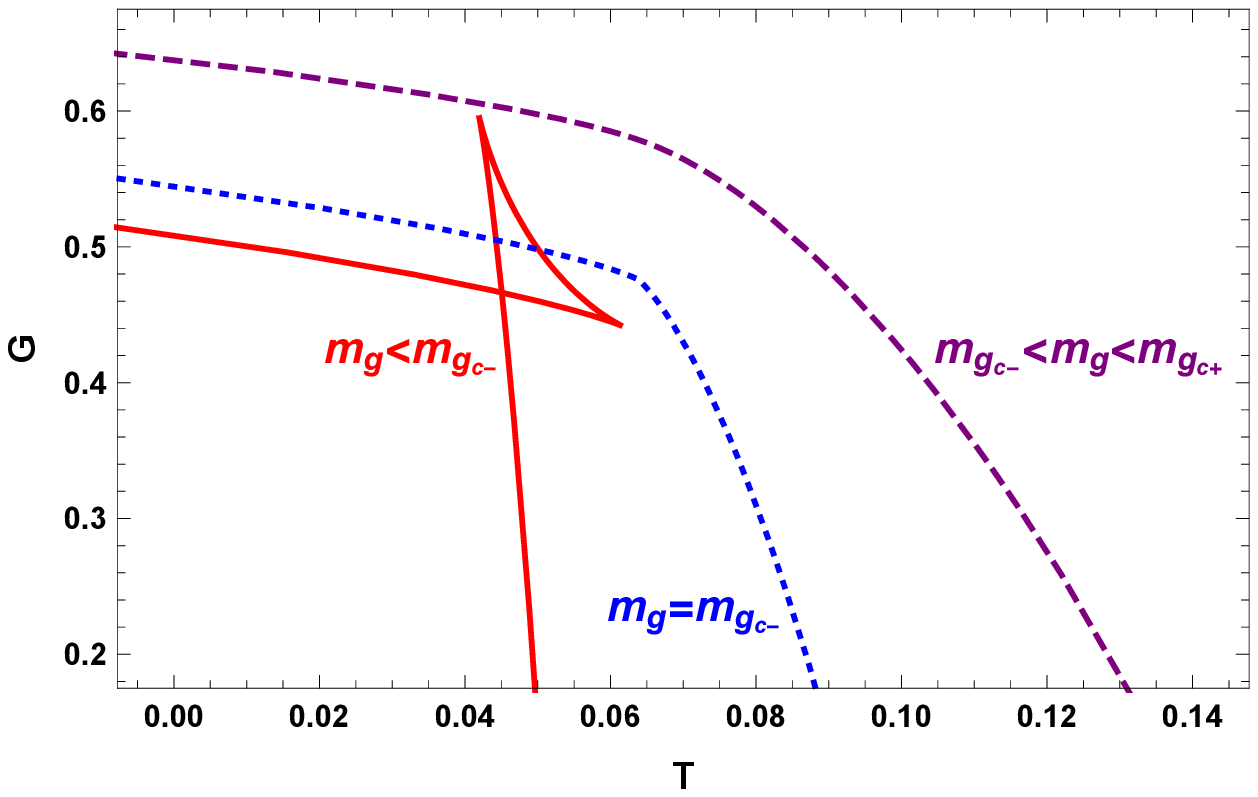} \>
			\includegraphics[width=7cm,height=4cm]{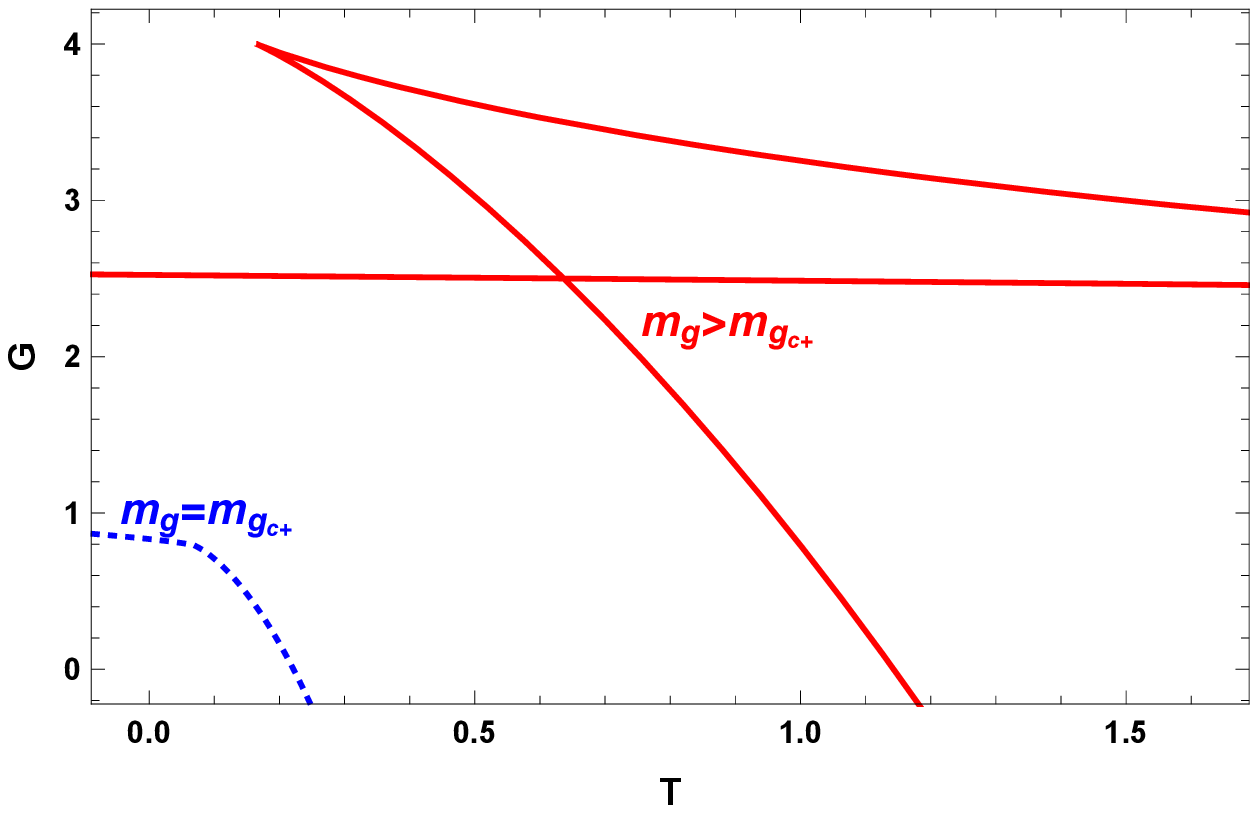} 
			
		\end{tabbing} 
	\caption{Canonical ensemble: The charged BH temperature as a function of the entropy. The two last panels shows the Gibbs Free energy as function of Hawking temperature, with parameter set $Q=0.5,\ c=1,\ \alpha=1\ \text{and}\ \beta=2$.}
	\label{figtherm2}
\end{center}
\end{figure}

 In Fig.~\ref{figtherm2}, we have plotted the temperature given by Eq.~\eqref{temperature} in terms of the entropy by setting $Q$ and $m_g$ to non-zero values. Like a first order phase transition of a van der Waals liquid, the temperature is not a monotonic function of $S$. However, we get two critical points $S_{1}$ and $S_{2}$ corresponding to the local maximum and minimum respectively, they are given by
\begin{eqnarray*}
	S_{1}&=&\pi \frac{c^2 \alpha ' m_g^2-\sqrt{\left(c^2 \alpha '
			m_g^2+1\right){}^2-36 Q^2 \gamma ' m_g^2}+1}{6 \gamma '
		m_g^2},\\
	S_{2}&=&\pi \frac{c^2 \alpha ' m_g^2+\sqrt{\left(c^2 \alpha '
			m_g^2+1\right){}^2-36 Q^2 \gamma ' m_g^2}+1}{6 \gamma '
		m_g^2}.
\end{eqnarray*}

The critical points located at $S_{1}$ and $S_{2}$  coalesce into a double point located at $S_{c\pm}$ resulting from the conditions
\begin{equation}\label{critique}
\left.\frac{\partial T}{\partial S}\right|_{m_{g_{c\pm}}}=0,\quad \left.\frac{\partial^2 T}{\partial S^2}\right|_{m_{g_{c\pm}}}=0.
\end{equation}

Using Eq.~\eqref{critique} and Eq.~\eqref{temperature} and assuming that $9 Q^2 \alpha^{\prime}>c^2 \gamma^{\prime}$, we get two critical points $(S_{c-},m_{g_{c-}})\ \text{and}\ (S_{c+},m_{g_{c+}})$ as follows
\begin{eqnarray}
S_{c\pm}&= &\pm \pi  Q \sqrt{9 Q^2-\frac{c^2 \alpha '}{\gamma '}}+3 \pi  Q^2,\\
m_{g_{c\pm}}&=&\frac{1}{\sqrt{9 Q^2 \gamma^{\prime}-c^2 \alpha^\prime}\pm 3 Q \sqrt{\gamma^{\prime}}}.
\end{eqnarray}
In the canonical ensemble, the  expressions of the free energy of the system, $G=M-T S$, read as,
\begin{equation}\label{gibbs}
G=\frac{\pi  S \left(c^2 m_g^2 \alpha^\prime+1\right)-m_g^2 S^2 \gamma^{\prime}+3 \pi ^2 Q^2}{4 \pi ^{3/2} \sqrt{S}}.
\end{equation}

In Fig.~\ref{figtherm2}, we plot $T$ versus $S$ and the function $G$ as a function of the temperature for different values of $m_g$, it can be seen that the characteristic "swallow tail" behaviour of $G –T$ curve appears for the first order phase transition where $m_g$ satisfying $m_g<m_{g_{c-}}$ and $m_g>m_{g_{c+}}$. However, within the range $m_{g_{c-}}<m_g<m_{g_{c+}}$  the swallow tail characteristic  disappears. At $m_g=m_{g_{c+}}$ and $m_g=m_{g_{c-}}$, we have the second order phase transition.

\begin{figure}[!h]
		\begin{center}
			\begin{tabbing}
				\hspace{8cm}\=\kill
		\includegraphics[width=7cm,height=4cm]{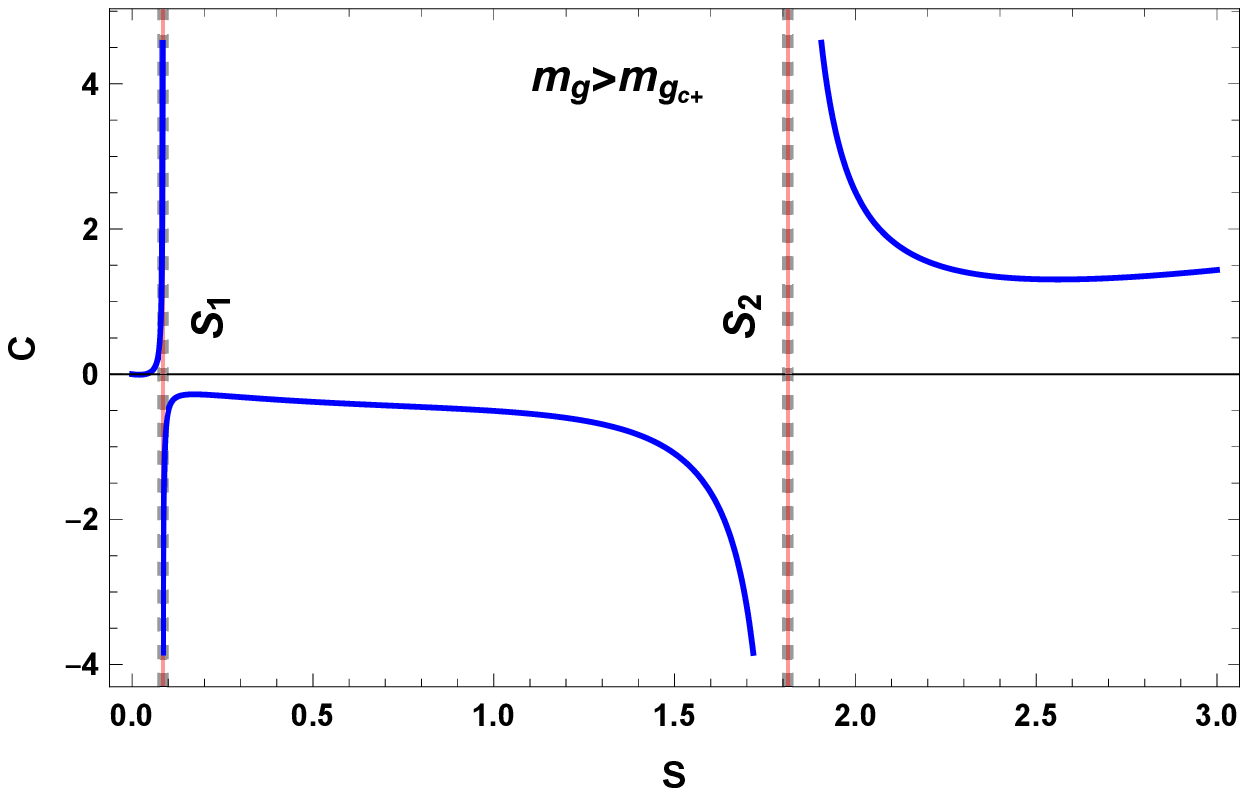} \>
		\includegraphics[width=7cm,height=4cm]{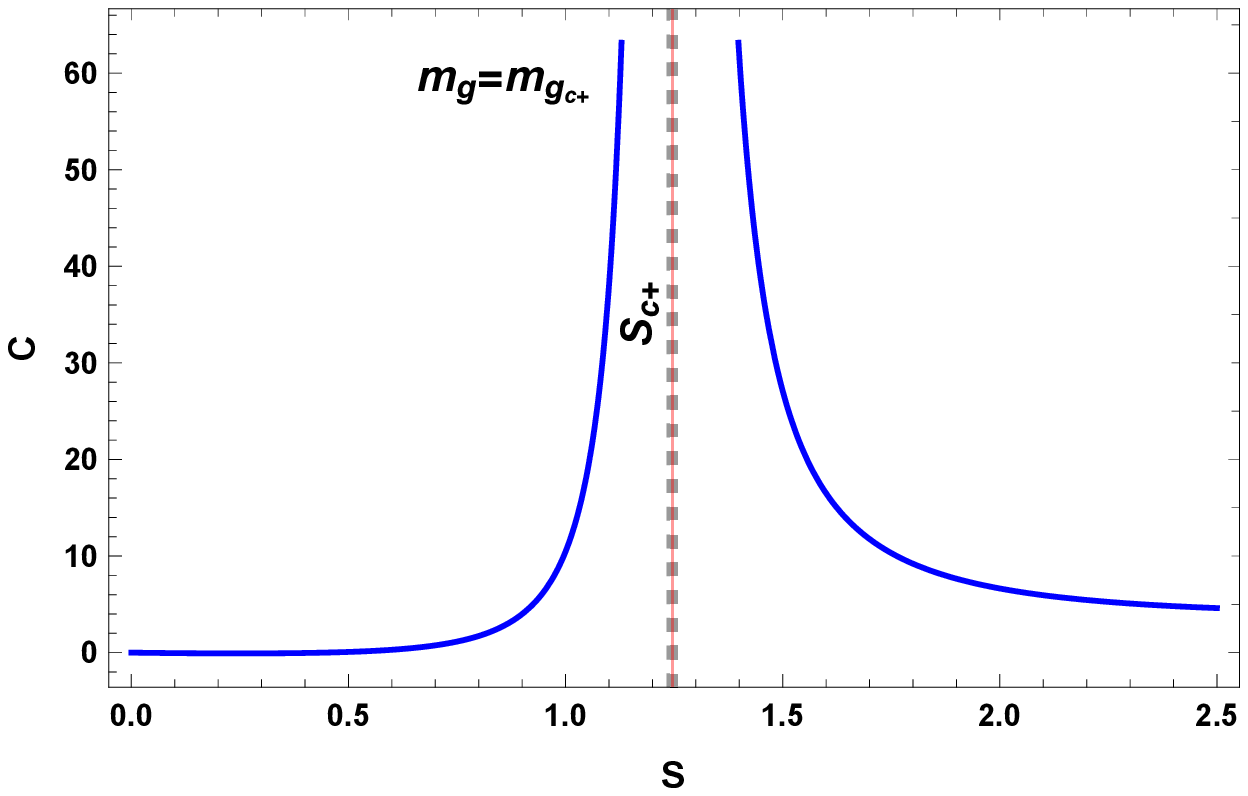} \\
		\includegraphics[width=7cm,height=4cm]{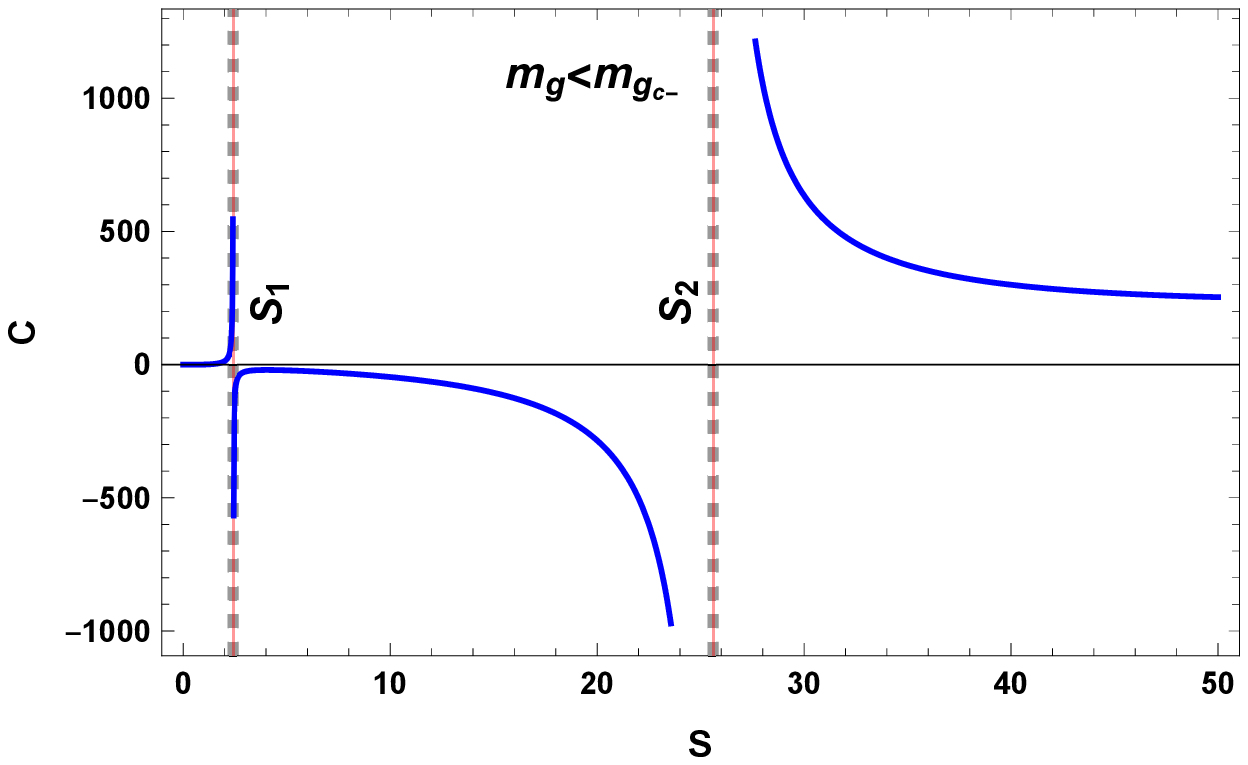} \>
		\includegraphics[width=7cm,height=4cm]{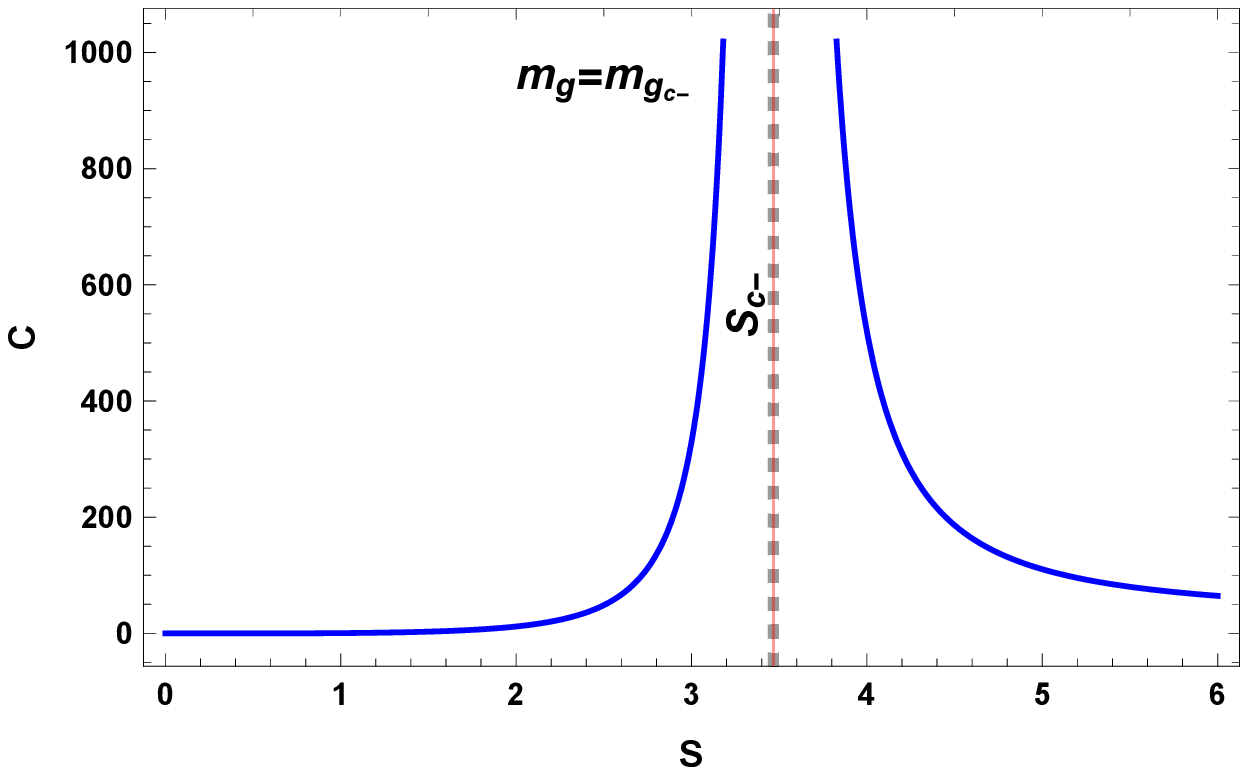} 
			\end{tabbing} 
	\caption{Canonical ensemble: The heat capacity as a function of the entropy for the charged BH where $m_g>m_{g_{c+}}$, there are two divergent points, with $Q=0.5,\ c=1,\ \alpha=1\ \text{and}\ \beta=2$. The dashed lines correspond to the divergency points of $C_{Q,m_g}$, the red lines correspond to the critical points.}
	\label{fig7}
\end{center}
\end{figure}

In this framework, the heat capacity $C$ at fixed $Q$  and $m_g$ can be written as,
\begin{equation} \label{3.1}
C_{Q,m_g}=\frac{2 S \left(\pi  S \left(c^2 m_g^2 \alpha^\prime+1\right)-2 \sqrt{\pi } c
   m_g^2 S^{3/2} \beta^{\prime}+3 m_g^2 S^2 \gamma^{\prime}-\pi ^2
   Q^2\right)}{-\pi  S \left(c^2 m_g^2 \alpha^\prime+1\right)+3 m_g^2
   S^2 \gamma^{\prime}+3 \pi ^2 Q^2}.
\end{equation}
We examine  Eq.~\eqref{3.1} for $m_g<m_{g_{c-}}$ and $m_g>m_{g_{c+}}$, we find that the denominator of $C_{Q,m_g}$ has two real roots, as shown in Fig.~ \ref{fig7}, one can check that the two critical points $S_{1}$ and $S_{2}$ correspond to the singularities of $C_{Q,m_g}$ (left panel of figure \ref{fig7}). For $m_g=m_{g_{c+}}$ and $m_g=m_{g_{c-}}$ we only have one divergence point at $S_c$ (right panel of Fig.~\ref{fig7}). We then conclude that the black hole is thermodynamically unstable for $S_{1}<S<S_{2}$ where he heat capacity is negative, while its thermodynamical stability is guaranteed  for $S<S_{1}$ and $S>S_{2}$, where $C_{Q,m_g}$ becomes positive.

\subsection {The grand canonical ensemble}
\begin{figure}[!h]
		\begin{center}
		\begin{tabbing}
			\hspace{8cm}\=\kill
		\includegraphics[width=7cm,height=4cm]{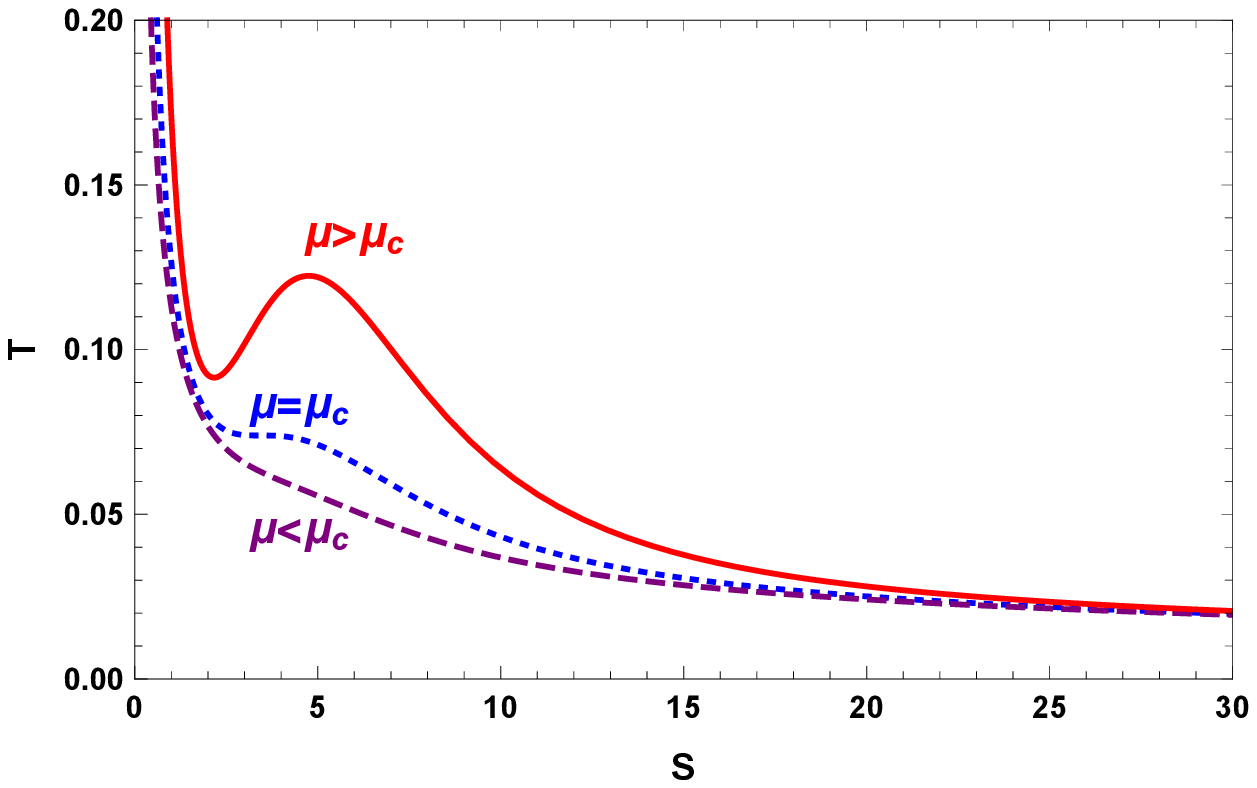} 	\> \includegraphics[width=7cm,height=4cm]{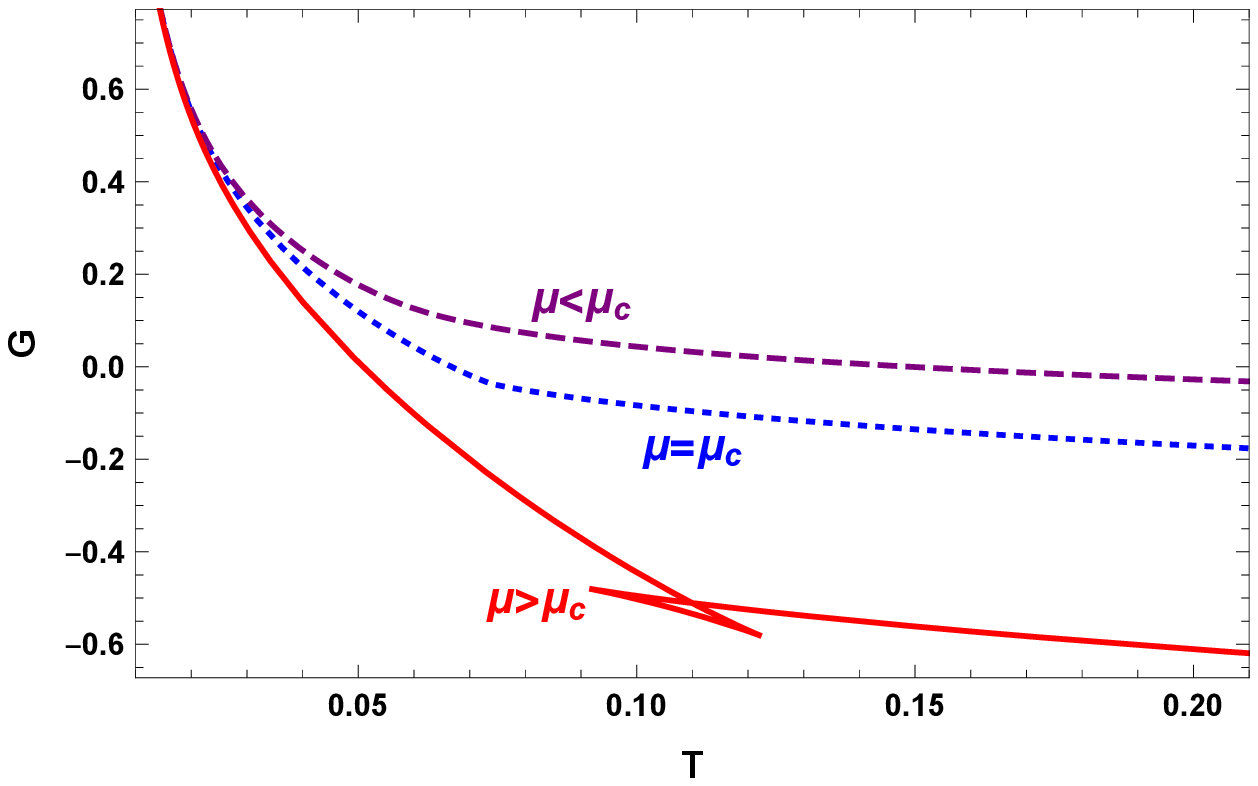} 
		\end{tabbing} 	
	\caption{Grand canonical ensemble: (left) Temperature as a function of $S$. (right) The Gibbs free energy versus $T$ for the neutral BH. We used the following values: $\Phi=0.5,\ c=1,\ \alpha=1\ \text{and}\ \beta=2$}
	\label{GGCC}
\end{center}
\end{figure}

In the grand canonical ensemble, besides the graviton potential, the electric potential is fixed as well. Thus, the Hawking temperature is given by,
\begin{equation}
T=\frac{\pi ^2 \mu ^2 \left(\pi  c^2 \alpha '-2 \sqrt{\pi } c \sqrt{S} \beta '+3 S \gamma '\right)+S
	\left(1-\phi ^2\right) \left(\pi  c^2 \alpha '-\sqrt{\pi } c \sqrt{S} \beta '+S \gamma '\right)^2}{4
	\sqrt{\pi } S^{3/2} \left(\pi  c^2 \alpha '-\sqrt{\pi } c \sqrt{S} \beta '+S \gamma '\right)^2}.
\end{equation}
Using Eq.~\eqref{massexp}, Eq.~\eqref{temperature}, Eq.~\eqref{chimicalpot} and Eq.~\eqref{chimique}, the Gibbs free energy, $G=M-T S-\Phi Q-\mu m_{g}$, can be derived in a straightforward way. In figure \ref{GGCC}, we plot $T$ versus $S$ (left panel) and $G$ versus $T$ (right panel).  The main outcome delivered by these  plots is that both charged and neutral black holes  have similar thermodynamical behaviour.

\begin{figure}[!h]
		\begin{center}
		\begin{tabbing}
			\hspace{8cm}\=\kill
			\includegraphics[width=7cm,height=4cm]{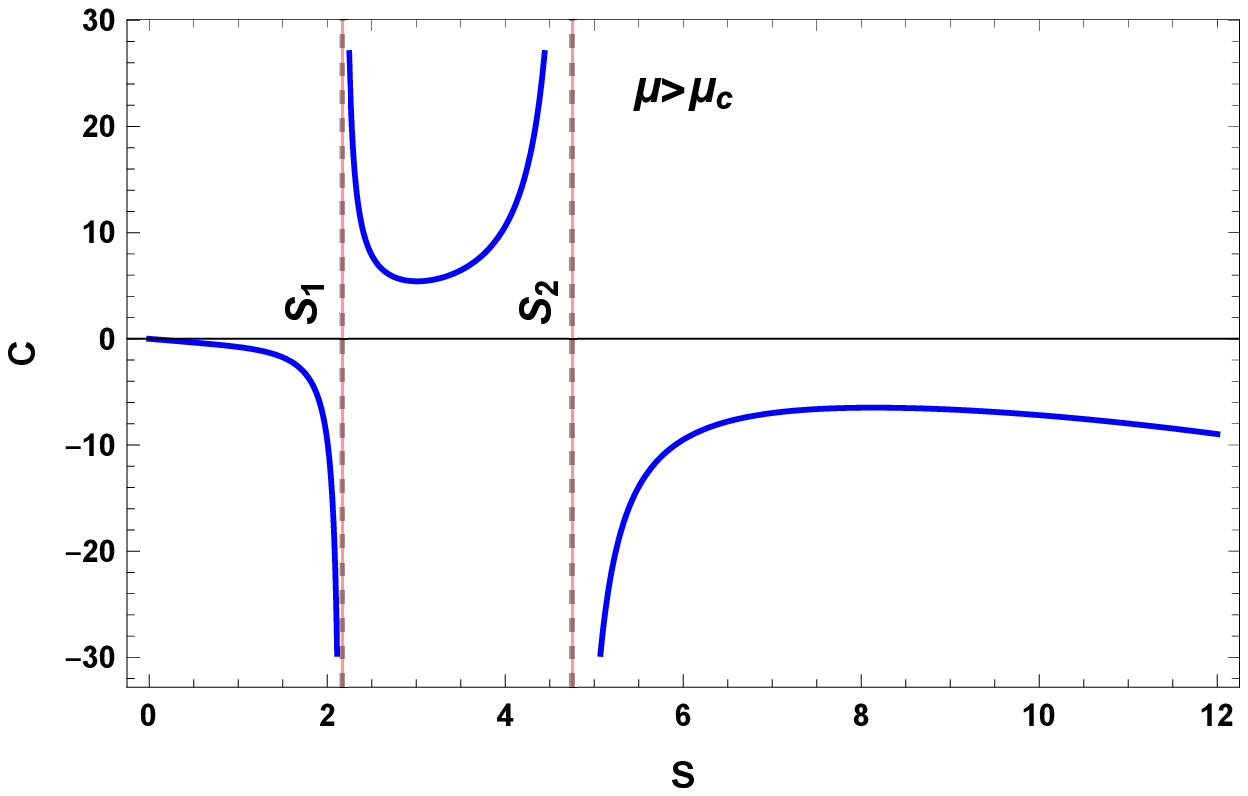} 	\>	\includegraphics[width=7cm,height=4cm]{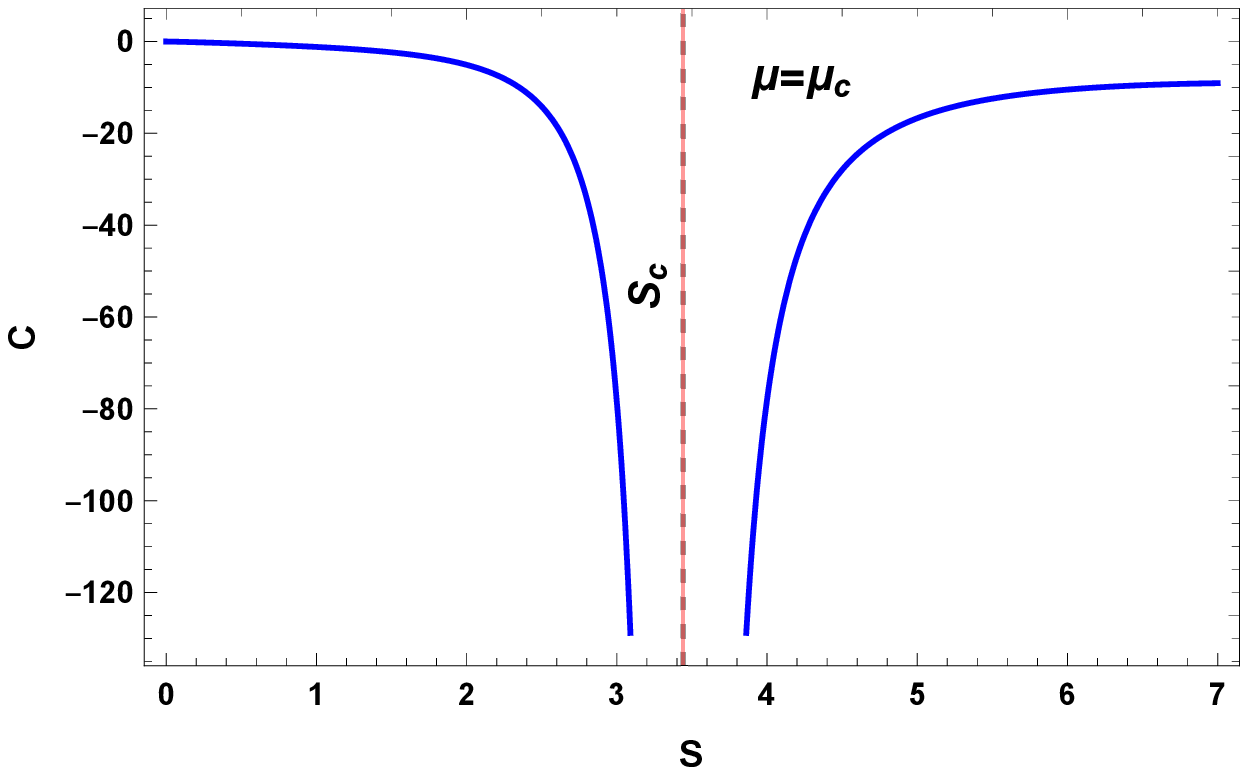}
		\end{tabbing} 		
	\caption{Grand Canonical ensemble: The heat capacity as a function of $S$, left panel for $\mu>\mu_{c}$ and the right panel is for $\mu=\mu_{c}$. The dashed lines correspond to the divergency points of $C_{\Phi,\mu}$, the red lines correspond to critical points. We used the following values:  $\Phi=0.5,\ c=1,\ \alpha=1\ \text{and}\ \beta=2$.} 
	\label{figC}
\end{center}
\end{figure}

 Besides, we also analyse the behaviour of the heat capacity $C_{\Phi,\mu}$ as a function of the entropy $S$ at constant $\mu$ and $\Phi$ in figure \ref{figC}. From the left panel, we see that $C_{\Phi,\mu}$ is negative for small black holes ($S<S_1$) and as well as for large black holes ($S>S_2$), while the heat capacity turns positive when $S_{1}<S<S_{2}$. Hence, one can conclude that the heat capacity diverges at the critical points $S_1$ and $S_2$,  revealing a first order phase transition, whereas the black hole undergoes a second order phase transition at $S_c$, as  illustrated by the right panel.
\FloatBarrier

\section{ Geometrical thermodynamics in dRGT massive gravity}\label{sec5}
In this section, we focus on the thermodynamical geometry of the black hole in dRGT massive gravity. Our objective is to check  whether the thermodynamical curvature encodes the singularities of the heat capacities and  consequently the critical behaviour of the black hole. To this end, two different geothermodynamics approaches are used: the first one, dubbed HPEM, has been proposed in \cite{Newmetric} and the second one is the free energy metric introduced in \cite{Liu2010sz}.

\subsection{HPEM metric}
The HPEM  metric is given by:

\begin{equation}
g^{HPEM}=S \frac{M_S}{M^{3}_{m_g m_g} M^{3}_{Q Q}}  \left(\begin{array}{ccc}-M_{SS} & 0&0 \\0& M_{QQ}&0\\0&0& M_{m_g m_g}\end{array}\right),
\end{equation}
where the total mass  is considered as the thermodynamical potential, while the charge and the graviton mass are treated as extensive parameters. 
It is worth noting that the HEPM metric is free from some drawbacks of the Ruppeiner and Quevedo metrics. 

In order to uncover the geometrical behaviour of HPEM metric, we calculate the Ricci scalar:

\begin{equation}\label{HPEM}
R^{HPEM}= \frac{A^{HPEM}}{B^{HPEM}},
\end{equation}
The divergency points in $R^{HPEM}$ appear when,
\begin{equation}
B^{HPEM}=B_1^{2}\times B_2^{3}=0,
\end{equation}
with
\begin{eqnarray*}
B_1&=&S m_g^2 \left(3 S \gamma^{\prime}-\pi  c^2 \alpha^{\prime}\right)+\pi  \left(3 \pi Q^2-S\right),\\
B_2&=&\pi  \left(\pi  Q^2-S\right)-S m_g^2 \left(\pi  c^2 \alpha^{\prime}-2 \sqrt{\pi } c
\sqrt{S} \beta^{\prime}+3 S \gamma^{\prime}\right).
\end{eqnarray*}

\begin{figure}[!h]
		\begin{center}
		\begin{tabbing}
			\hspace{8cm}\=\kill
			\includegraphics[width=7cm,height=4cm]{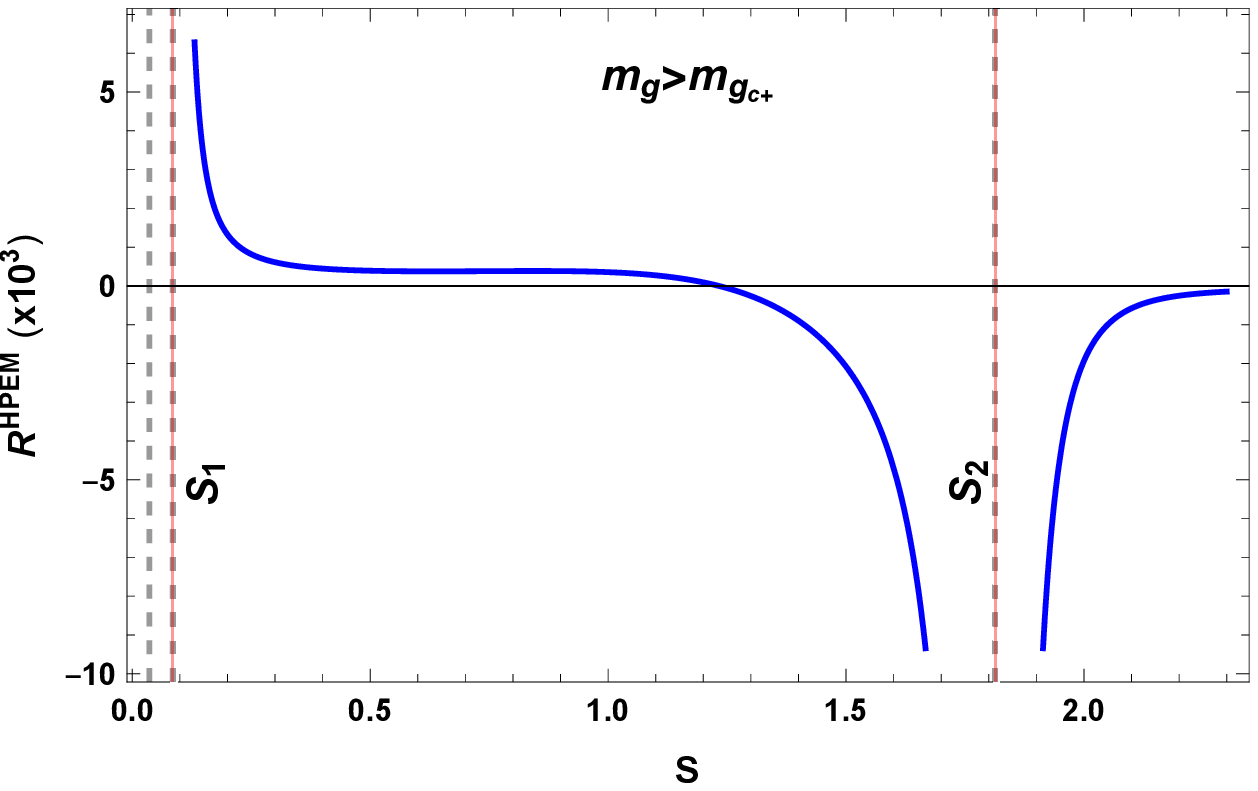}\>\includegraphics[width=7cm,height=4cm]{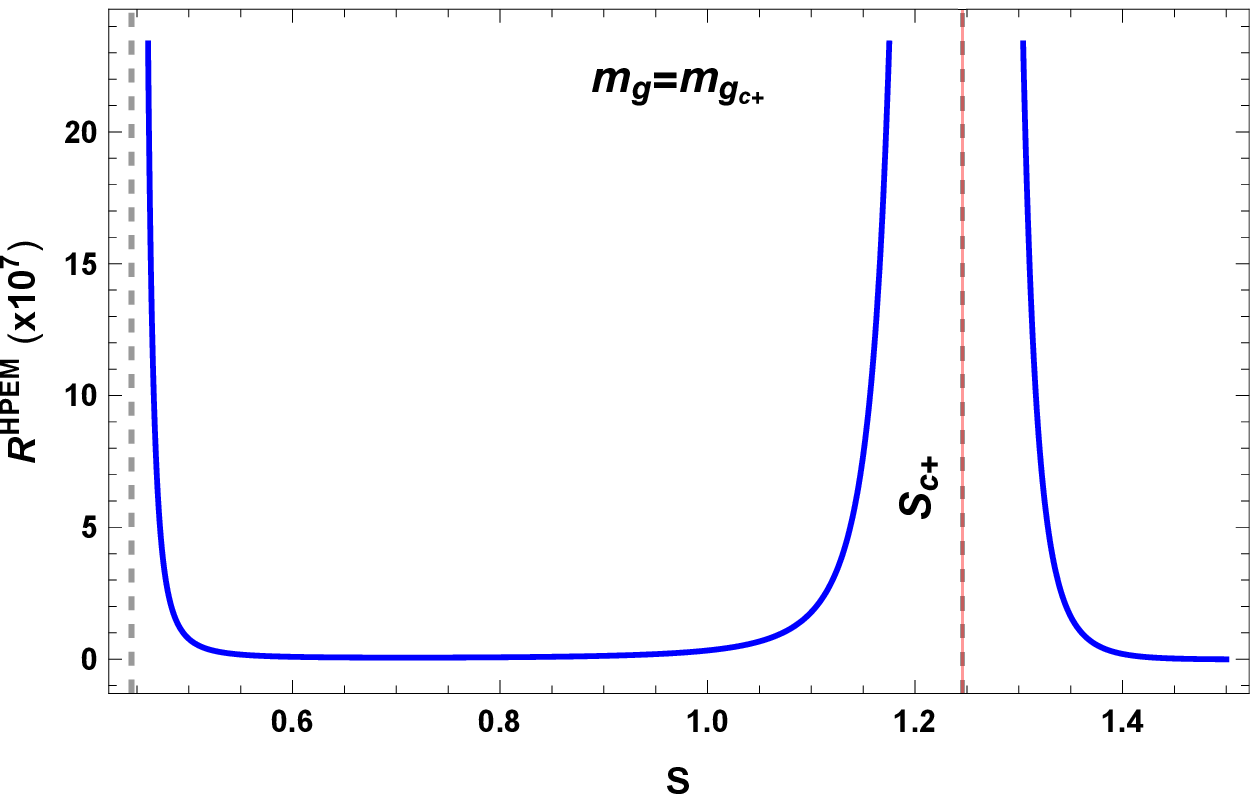}\\
			\includegraphics[width=7cm,height=4cm]{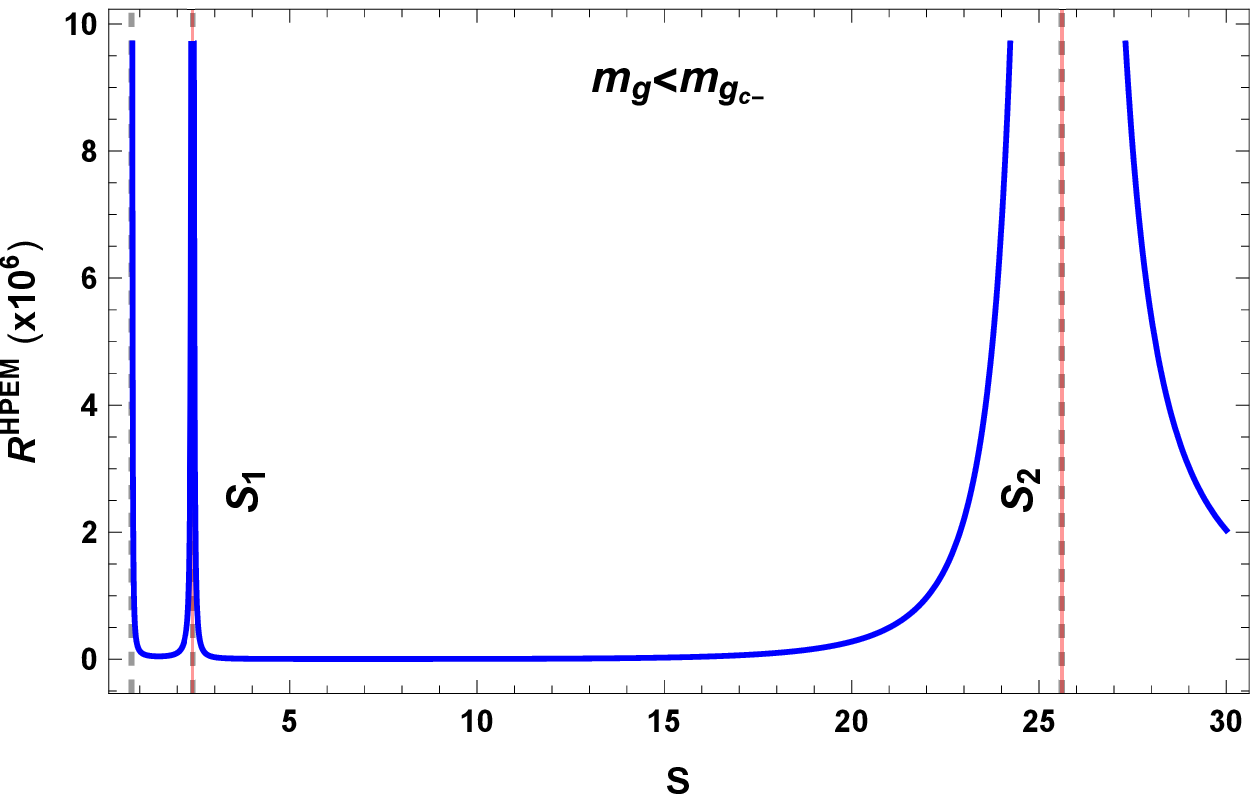}\>\includegraphics[width=7cm,height=4cm]{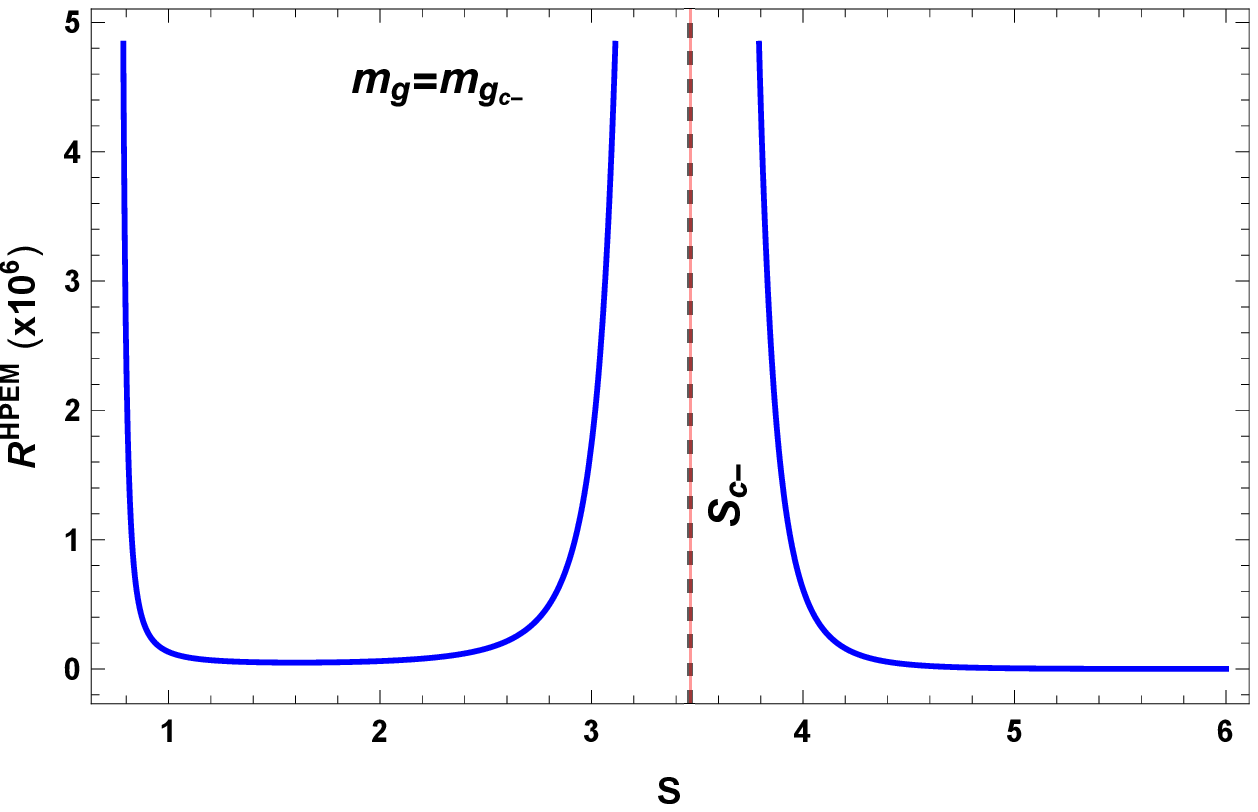}
		\end{tabbing}
		\caption{The scalar curvature versus the entropy for the HPEM metric geometry. The dashed lines correspond to the divergency points of $R^{HPEM}$ and  red lines to critical points. We used the following values: $Q=0.5$, $\alpha=1$, $\beta=2$ and $c=1$. }
		\label{fig9}
			\end{center}
	\end{figure}

Here, the results are more favorable than that for Quevedo, indeed the HPEM scalar curvature shares the same factor $B_1$ in the denominator as the specific heat does, while the expression of $B_2$ is same as the temperature in Eq.~\eqref{temperature}. Therefore the curvature $R^{HPEM}$ will exactly diverge at those points at which the heat capacity $C_{Q,m_g}$is singular as well as at the extremal black hole, as graphically illustrated  in Fig.~\ref{fig9}. 
Notice that $R^{HPEM}$ doesn't diverge in $S=0$ as can be assumed from Fig.~\ref{fig9}. Indeed from a simple check of Eq. (34), we can see that   the denominator of $R^{HPEM}$ reduces to $B=9 \pi ^{10} Q^{10}$ when the entropy becomes null.

\subsection{The free-energy metric}
Based on the Hessian matrix of several free energy generated by the Legendre transformations of $M$, we  consider the Gibbs free energy $G=M-T S$ as a function of $T$, $Q$ and $\mu$ \cite{Liu2010sz}.  In differential form, we get
\begin{equation}
dG=-SdT+\Phi dQ+\mu dm_{g}.
\end{equation}
Then, it follows that the Gibbs free energy metric is given by
\begin{equation}
ds_{G}^{2}=-dTdS+d\Phi dQ+d\mu dm_{g}.
\end{equation}
Within the natural variables $\left(T,Q,m_{g}\right)$, it is convenient here to trade the temperature by the entropy $S$. By using \eqref{temperature}, \eqref{chimicalpot} and \eqref{chimique},  the resulting metric becomes
\begin{equation}
ds_{G}^{2}=g_{SS}dS^{2}+g_{QQ}dQ^{2}+g_{m_{g}m_{g}}dm_{g}^{2}+2 g_{SQ}dSdQ+2 g_{Sm_{g}}dSdm_{g}+2 g_{Qm_g}dQdm_g,
\end{equation}
with
\begin{eqnarray}
g_{SQ}&=& g_{Sm_{g}}=g_{Qm_{g}}=0,\\
g_{SS}&=& \frac{S m_g^2 \left(\pi  c^2 \alpha '-3 S \gamma '\right)+\pi 
	\left(S-3 \pi  Q^2\right)}{8 \pi ^{3/2} S^{5/2}},\\
g_{QQ}&=& \sqrt{\frac{\pi}{S}},\\
g_{m_g m_g}&=& \pi^{-\frac{2}{3}} \left[\pi  c^2 \sqrt{S} \alpha '-\sqrt{\pi } c S \beta '+S^{3/2}
\gamma ' \right].
\end{eqnarray}
So, after a straightforward calculation,  the scalar curvature can be expressed as
\begin{equation}\label{Rgibbs}
R^{G}=\frac{A^{G}}{B_{1}^{2} \times B_{3}^{2}} ,
\end{equation}
with, 
\begin{eqnarray*}
	B_{1}&= &S m_g^2 \left(3 S \gamma '-\pi  c^2 \alpha '\right)+\pi  \left(3
	\pi  Q^2-S\right), \\
	B_{3}&= &\pi  c^2 \alpha '-\sqrt{\pi } c \sqrt{S} \beta '+S \gamma ' .
\end{eqnarray*}

\begin{figure}[!h]
	\begin{center}
		\begin{tabbing}
			\hspace{8cm}\=\kill
			\includegraphics[width=7cm,height=4cm]{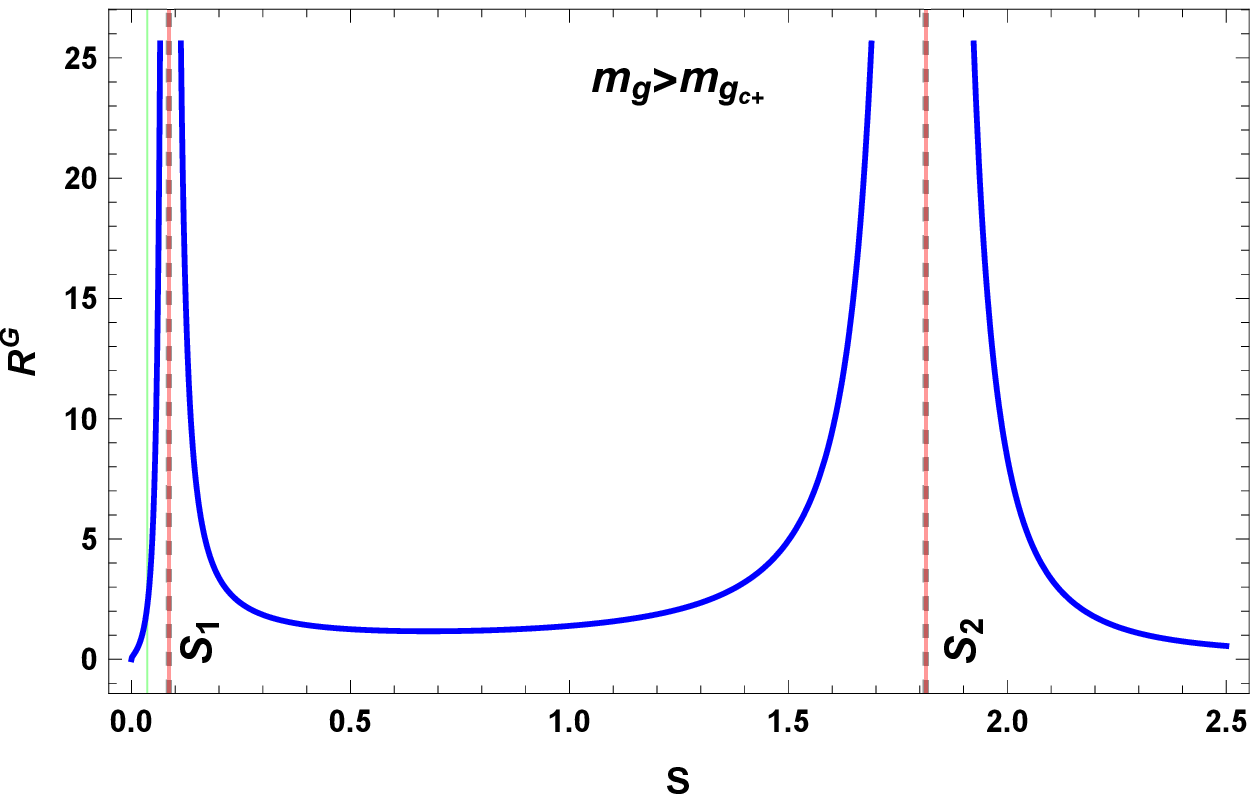}\>\includegraphics[width=7cm,height=4cm]{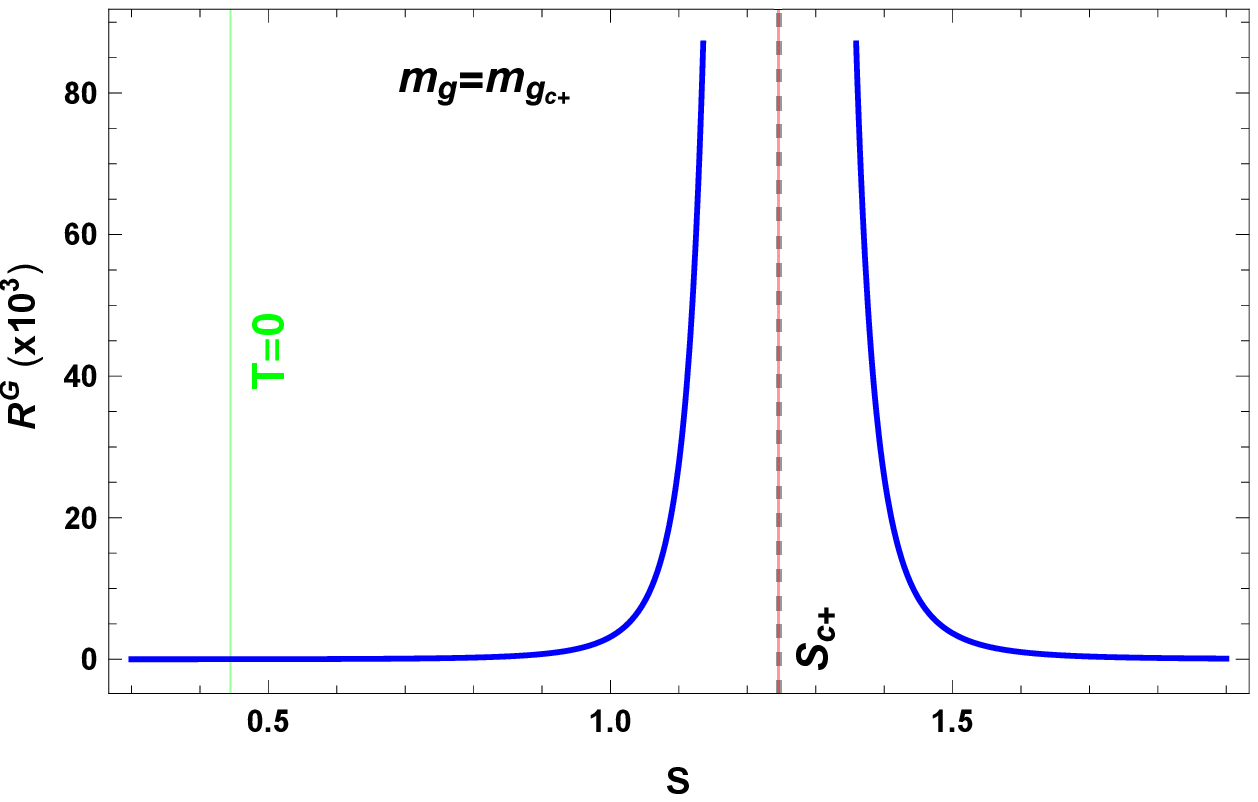}\\
			\includegraphics[width=7cm,height=4cm]{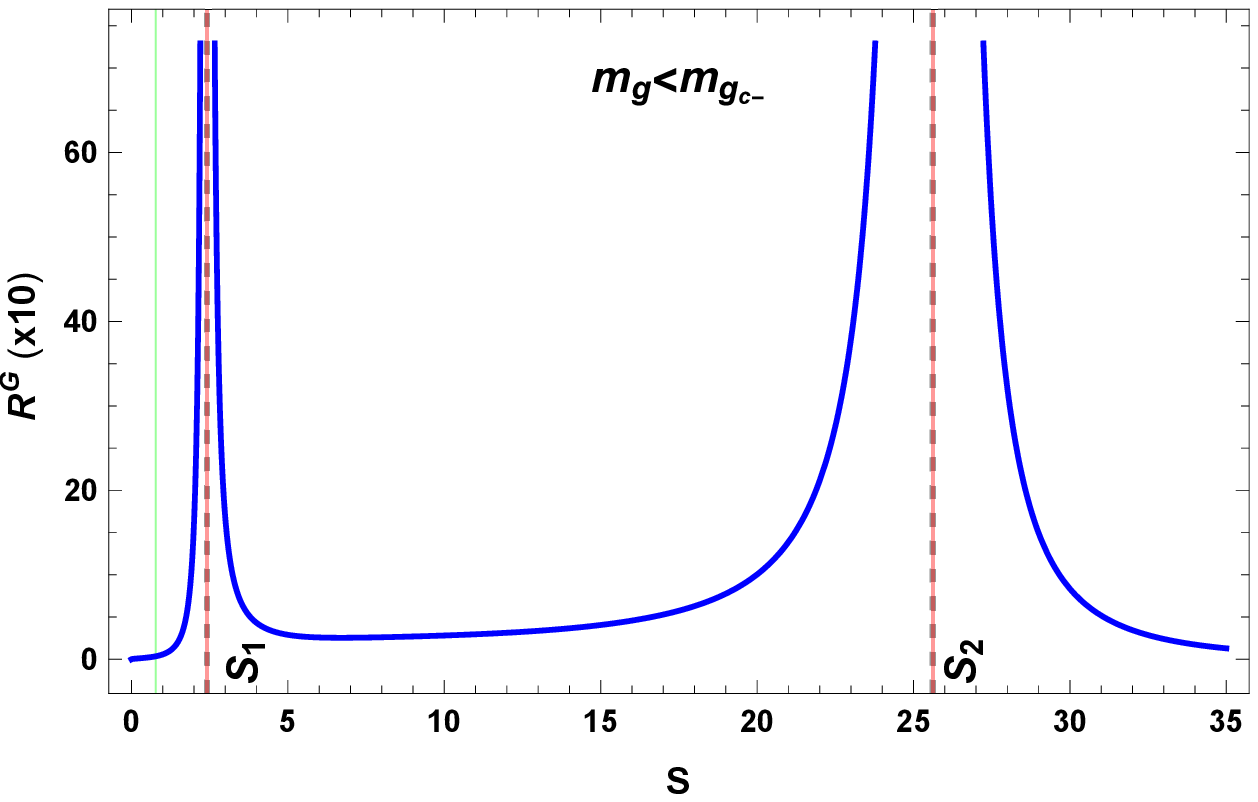}\>\includegraphics[width=7cm,height=4cm]{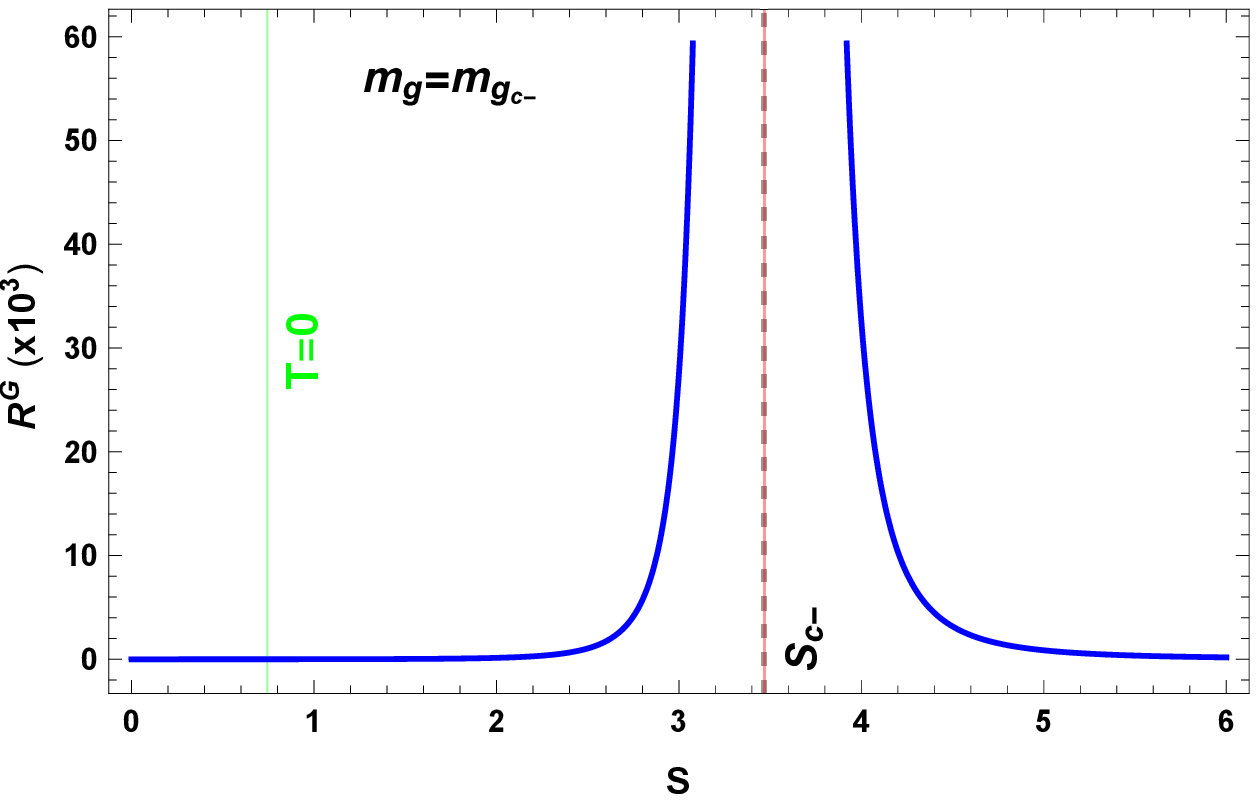}
		\end{tabbing}
		\caption{The scalar curvature versus the entropy for the geometry of the free energy metric. The dashed lines correspond to the divergency points of $R^{G}$, the red lines to critical points and green lines gives the extremal black hole where $T=0$. We used the following values: $Q=0.5$, $\alpha=1$, $\beta=2$ and $c=1$. }
		\label{fig10}
	\end{center}
\end{figure}

First, note that the Ricci scalars  $R^{G}$ Eq.~\eqref{Rgibbs} and $R^{HPEM}$ Eq.~\eqref{HPEM} share the same term $B_{1}$ in their respective denominator. Furthermore,  $B_{1}$  also appears in the denominator of the heat capacity as can be seen from Eq.~\eqref{3.1}. Hence, as illustrated in Fig.~ \ref{fig10},  the divergent points of the heat capacity are exactly located within the collection of the scalar curvature singularities  of the Gibbs free energy metric. However, comparatively to the HPEM metric,  $R^{G}$ is insensitive to the extremal black hole, and develops other singularities located at $B_{3}=0$ which do not coincide with those of the heat capacity $C_{Q,m_g}$.  However, these singularities correspond to the divergent points of the capacity $\tilde{C}_{Q,S}$ related to the graviton mass  at fixed charge $Q$ and entropy $S$, since $B_{3}$ shows up  in the $\tilde{C}_{Q,S}$ denominator, as can be seen from the following formula:
 \begin{equation}\label{Capacitygraviton}
\tilde{C}_{Q,S}=\left. \frac{\partial m_{g}}{\partial \mu}\right|_{Q,S} =\frac{\pi ^{3/2}}{\sqrt{S} \left(\pi  c^2 \alpha '-\sqrt{\pi } c \sqrt{S} \beta '+S \gamma ' \right)}.
 \end{equation}

To summarise, our results confirm the recent results obtained in \cite{67} which pointed out that the divergencies of the scalar curvature for the HPEM metric coincide with those of the heat capacity. Consequently, the HPEM metric might be considered as a powerful tool to describe the phase structures of black holes in dRGT massive gravity. As to the free energy  geometry,  we have seen that the analysis of the singular behavior of its Ricci scalar can well measure the black hole phase transition. However, $R^{G}$ is not sensitive to the extremal black hole and develops other divergent points which rather coincide with those of the other capacity $\tilde{C}_{Q,S}$. These features confirm that only the metric of an appropriate thermodynamical potential is relevant to a particular phase transition, as noticed in \cite{Liu2010sz}.

\FloatBarrier
\section{Conclusion}
In this paper, we have studied a black hole in dRGT massive gravity, where the graviton mass has been treated as a thermodynamic variable with the graviton potential as its conjugate quantity.  In this framework, we have determined the critical points and the analysed divergent behaviour of specific heat in the canonical and grand canonical ensemble. More precisely, we have examined the $T-S$ criticality using the standard thermodynamic techniques and revealed that the system has double critical points located at $(S_{c-},m_{g_{c-}})\ \text{and}\ (S_{c+},m_{g_{c+}})$. According to the value of $m_g$, we obtained three scenarios:
\begin{itemize}
	\item $m_g=m_{g_{c-}}$ and $mg=m_{g_{c+}}$: the black hole undergoes a second order phase transition;
	\item $m_g<m_{g_{c-}}$ and $mg>m_{g_{c+}}$:  the first order phase transition shows up;
	\item $m_{g_{c-}}<m_g<m_{g_{c+}}$, no phase transition.
\end{itemize}

Besides, we have also employed two geometric approaches to check the results found above for thermodynamics phase transitions. Here the graviton mass and the charge $Q$ have been chosen as an extensive variables for calculating the scalar curvatures. First, by exploring the singularities for the scalar curvature of HPEM metric, our analysis showed that the roots of all terms appearing in the scalar curvature denominator  exactly coincide with those of the heat capacity. Therefore, the obtained features on the phase structure of black holes in dRGT massive gravity are well established. Then, we constructed the thermodynamic geometry with the free energy metric  and calculated the corresponding scalar curvature. The analysis  have shown that the Ricci scalar  can well reveal divergent  behavior of the heat capacity, which means that the free energy geometry is an efficient tool to gain insight into phase transitions of black holes.

At last, we would like to mention that a variety of methods exist in literature to study black hole thermodynamics 
geometry.  As example, the main features of Ruppeiner and Quevedo metrics are briefly presented in the appendix.

\appendix
\section{Appendix}
\subsection{Ruppeiner phase space}\label{a1}
The Ruppeiner metric as defined by Ruppeiner \cite{hep49} has the following form
\begin{equation}\label{3.5}
ds_R^2=\frac{1}{T} \frac{\partial^{2} M}{\partial X^{\alpha} \partial X^{\beta}},
\end{equation}
where $X^{\mu}=\{S,Q,m_{g}\}$.  The Ruppeiner metric can be written as:
\begin{equation}\label{3.9}
g^{R}=\frac{1}{T}\left(\begin{array}{ccc}M_{SS} & M_{SQ} & M_{Sm_g} \\M_{Q S} & M_{QQ} & M_{Qm_g}\\M_{m_gS} & M_{m_gQ}& M_{m_gm_g}\end{array}\right).
\end{equation}

The resulting general form of the scalar curvature of this metric reads as:
\begin{equation}\label{scalarform}
R^{R}=\frac{A^{R}}{B^{R}},
\end{equation}
where $A^{R}$ and $B^{R}$ stand for the numerator and the denominator of $R^{R}$. We have:
\begin{equation}\label{rupdenom}
B^{R}=B_2\times B_4^{2},
\end{equation}
with,
\begin{eqnarray*}
	B_2&=&\pi  \left(\pi  Q^2-S\right)-S m_g^2 \left(\pi  c^2 \alpha^{\prime}-2 \sqrt{\pi } c
	\sqrt{S} \beta^{\prime}+3 S \gamma^{\prime}\right),\\	
B_4&=&	\pi  c \left(\sqrt{\pi } \sqrt{S} \beta ' \left(-9 c^2 S \alpha '
	m_g^2+\pi  Q^2-S\right)+\pi  c \alpha ' \left(3 c^2 S \alpha '
	m_g^2-\pi  Q^2+S\right)+8 c S^2 \left(\beta '\right)^2
	m_g^2\right) \\
	& & + \pi  S \gamma ' \left(10 c^2 S \alpha ' m_g^2-\pi 
	Q^2+S\right)-21 \sqrt{\pi } c S^{5/2} \beta ' \gamma ' m_g^2+15
	S^3  (\gamma')^{2} m_g^2.  \nonumber	
\end{eqnarray*}
Nota that $B_2\propto T$, while $B_{4}$ is not identical with the denominator of  the heat capacity. Hence the singularities of scalar curvature may not coincide with the divergencies of the heat capacity.

\subsection{Quevedo phase space}\label{a2}
To have a Legendre invariant geometry  in the space of equilibrium state and to cure the failures of Ruppeiner metrics \cite{hep49}, Quevedo proposed a new thermodynamics metric \cite{hep54} given by: 
\begin{equation}
g^{Quv}=(S T+\mu\cdot m_g+Q \Phi) \left(\begin{array}{ccc}-M_{SS} & 0 & 0 \\0&M_{Q Q}& M_{Qm_g}\\0&M_{m_g Q}&M_{m_g m_g}\end{array}\right),
\end{equation}

Here, the resulting scalar curvature can be written as:
\begin{equation}\label{scalarformQ}
R^{Quv}=\frac{A^{Quv}}{B^{Quv}},
\end{equation}
with,
\begin{equation}
B^{Quv}=B_{1}^{2}\times B_{3}^{2}\times  B_{5}^{3},
\end{equation}
and
\begin{eqnarray*}
	B_1&=&S m_g^2 \left(3 S \gamma^{\prime}-\pi  c^2 \alpha^{\prime}\right)+\pi  \left(3 \pi Q^2-S\right),\\
	B_3&=&\pi  c^2 \alpha^{\prime}-\sqrt{\pi } c \sqrt{S} \beta^{\prime}+S \gamma^{\prime}, \\
	B_5&=& S m_g^2 \left(5 \pi  c^2 \alpha^{\prime}-6 \sqrt{\pi } c \sqrt{S} \beta^{\prime}+7 S \gamma^{\prime}\right)+\pi  \left(3 \pi  Q^2+S\right).
\end{eqnarray*}

We clearly see  that the second part, $B_{1}$, of this denominator is identical to the denominator appearing in the heat capacity. Therefore, the curvature will diverge exactly at singular point of the heat capacity, while $B_3$ and $B_5$ terms in $R^{Quv}$ develop other  divergent points that do not coincide with the specific points of $C_{Q,m_g}$.

\end{document}